\pdfoutput=1
\documentclass[fleqn,usenatbib]{mnras}

\usepackage{newtxtext,newtxmath}

\usepackage[T1]{fontenc}

\DeclareRobustCommand{\VAN}[3]{#2}
\let\VANthebibliography\thebibliography
\def\thebibliography{\DeclareRobustCommand{\VAN}[3]{##3}\VANthebibliography}

\usepackage{graphicx}	
\usepackage{amsmath}	
\usepackage{float}

\usepackage{hyperref}
\usepackage{caption}
\usepackage{subcaption}
\usepackage[output-decimal-marker={,}]{siunitx}
\usepackage{comment}
\usepackage{gensymb}

\title[Transition Region Oscillations]{Spectroscopic study of solar transition region oscillations in the quiet-Sun observed by IRIS using Si IV spectral line}

\author[Sangal et al.]{
Kartika Sangal,$^{1}$
A.K.~Srivastava,$^{1}$\thanks{E-mail: asrivastava.app@iitbhu.ac.in}
P.~Kayshap,$^{2}$
T.J~Wang$^{3,4}$
J.J. Gonz\'{a}lez-Avil\'{e}s,$^{5}$
and Abhinav Prasad$^{1}$
\\
$^{1}$Department of Physics, Indian Institute of Technology (BHU), Varanasi-221005, India\\
$^{2}$Vellore Institute of Technology, Kotri Kalan, Ashta, Near, Indore Road, Bhopal, Madhya Pradesh 466114\\
$^{3}$The Catholic University of American (CUA), USA\\
$^{4}$NASA Goddard Space Flight Center, Code 671, 20771, Greenbelt, MD, USA\\
$^{5}$Investigadores por M\'exico-Conacyt, Servicio de Clima Espacial M\'exico, Laboratorio Nacional de Clima Espacial, Instituto de Geof\'{i}sica, \\ Unidad Michoac\'{a}n, Universidad Nacional Aut\'{o}noma de M\'{e}xico, 58190 Morelia, Michoac\'{a}n, M\'{e}xico\\
}

\date{Accepted XXX. Received YYY; in original form ZZZ}

\pubyear{2021}

\begin{document}
\newcommand{\Int}{\int\limits}
\label{firstpage}
\pagerange{\pageref{firstpage}--\pageref{lastpage}}
\maketitle
\begin{abstract}


In the present paper, we use Si IV 1393.755 \AA\ spectral line observed by the Interface Region Imaging Spectrograph (IRIS) in the quiet-Sun to determine physical nature of the solar transition region (TR) oscillations. We analyze the properties of these oscillations using wavelet tools (e.g., power, cross-power, coherence, and phase difference) along with the stringent noise model (i.e., power-law + constant). We estimate the period of the intensity and Doppler velocity oscillations at each chosen location in the quiet-Sun (QS) and quantify the distribution of the statistically significant power and associated periods in one bright and two dark regions. In the bright TR region, the mean periods in intensity and velocity are 7 min, and 8 min respectively. In the dark region, the mean periods in intensity and velocity are 7 min, and 5.4 min respectively. We also estimate the phase difference between the intensity and Doppler velocity oscillations at each location. The statistical distribution of phase difference is estimated, which peaks at -119\degree $\pm$ 13\degree, 33\degree $\pm$ 10\degree, 102\degree $\pm$ 10\degree\ in the bright region, while at -153\degree $\pm$ 13\degree, 6\degree $\pm$ 20\degree, 151\degree $\pm$ 10\degree\ in the dark region. The statistical distribution reveals that the oscillations are caused by propagating slow magnetoacoustic waves encountered with the TR. Some of these locations may also be associated with the standing slow waves. Even, in the given time domain, several locations exhibit presence of both propagating and standing oscillations at different frequencies.

\end{abstract}


\section{Introduction}

The solar atmosphere exhibits waves and oscillations which may be observed in different layers of the solar atmosphere, i.e. photosphere, chromosphere, transition region (TR), and inner corona \citep[e.g.,][]{2001ApJ...554..424J,2003ApJ...595L..63D,2007A&A...474..627Z,2007A&A...467.1299E,2009A&A...494..355O,2012ApJ...757..160J,2018MNRAS.479.5512K,2020A&A...634A..63K}. In different parts of the Sun's outer atmosphere, various periods of oscillation have been observed \citep[e.g.,][]{1999ApJ...511L.121B,2001A&A...371.1137B,2002SoPh..207..259B,2008A&A...488..331T,2009SSRv..149...65D,2012SoPh..281...67K,2018ApJ...855...65H}. The oscillations in the solar atmosphere are typically explained in terms of various magnetohydrodynamic modes \citep[e.g,][]{2000A&A...355L..23D,2002A&A...387L..13D,2006A&A...452.1059O,2010ApJ...713..573M,2015ApJ...812L..15K,2022ApJ...924..100C}. 

In the solar atmosphere, there are multitudes of waves including magneto-acoustic waves and Alfv\'en waves \cite[e.g.,][and references therein]{2015SSRv..190..103J,2016GMS...216..395W,2020ARA&A..58..441N,2021SSRv..217...34W,2021SSRv..217...76B}.
These wave modes can carry energy throughout the different layers of the solar environment. Wave propagation is an efficient way of transporting energy between layers of the solar atmosphere, and waves can contribute significantly to the heating of the chromosphere and corona \cite[e.g.,][and references therein]{2003A&ARv..12....1W,2006A&A...453.1067K,2006ApJ...648L.151J,2009SSRv..149..229T,2015RSPTA.37340261A,2020A&A...642A..52A,2020SSRv..216..140V,2020JApA...41...18Z,2021A&A...648A..28A,2021JGRA..12629097S}. The theory underlying MHD waves was established several decades ago, and its applications are becoming more important as high-resolution observational data now become available.

The TR is located between the chromosphere and the corona and is defined by temperature variation rather than height variation. The temperature ranges from 20,000 to 1,000,000 K in the TR and it is a highly complex, magnetically structured region \citep[e.g.,][and references therein]{1992str..book.....M,2010NewAR..54...13T,2017RAA....17..110T}. Our understanding of the TR is enriched by observations made by the Solar Ultraviolet Measurements of Emitted Radiation (SUMER) \citep{1995SoPh..162..189W} onboard the Solar and Heliospheric Observatory (SOHO) and the Interface Region Imaging Spectrograph (IRIS) \citep[e,g.,][]{2014SoPh..289.2733D,2021SoPh..296...84D}. In the TR, different periods of oscillation have been reported. For example, \citet{2006A&A...448.1169G} studied the oscillations in the intensity and Doppler shift above the network region in a TR spectral line and reported a period in the range of 250-450 sec. They have proposed the signature of downward propagating waves in the TR. \cite{1998SoPh..181...51D} reported a period of oscillations in the intensity in the range of 200-500 sec in TR. \cite{2009A&A...493..217T} reported 5 min oscillations in the dark mottle and network boundaries, and suggested the upward propagating waves at the network boundary, and standing waves at the mottle region. In the previous studies, 3 min oscillations are widely reported in the TR spectral line formed above the sunspots, and interpreted as a signature of upward propagating waves \citep[e.g.,][]{1982ApJ...253..939G,1999ApJ...511L.121B,2001A&A...373L...1M,2001A&A...368..639F,2002A&A...387..642O,2003A&A...398L..15B,2004SoPh..221..237B,2005A&A...444..585L,2012A&A...539A..23S}.
\cite{2008ApJ...682.1363J} studied the velocity oscillations in the bright active region at the TR height and reported high frequency oscillations with a dominant period of 26 $\pm\ 4$ sec. They have interpreted these oscillations as the signature of fast global sausage modes. \citet{2003ApJ...595L..63D} studied the oscillations in the TR above the plage and found the oscillations with periods ranging from 200 to 600 sec.
\cite{2000SoPh..196...63B} have reported long period oscillations (above 10 min) in the TR spectral line, and suggested that these oscillations are essentially the propagating slow magneto-acoustic waves \citep[e.g., see also][]{ 2001A&A...377..691B,2001A&A...380L..39B}. \\


\begin{figure*}
\includegraphics[height=8cm,width=18cm]{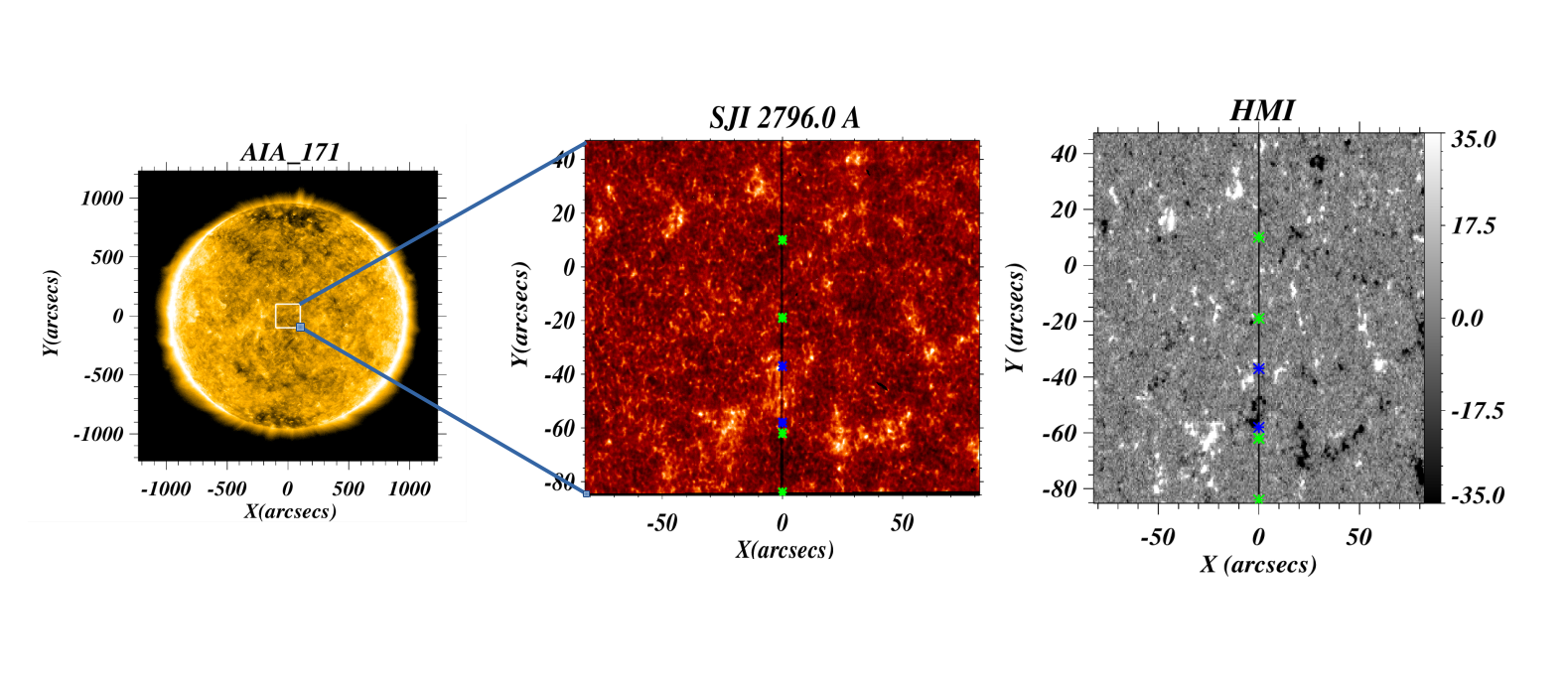}
   \caption{The left panel shows the full context image of the Sun as observed by SDO/AIA in 171 \AA\ passband. The overlaid white box depicts the region of interest. The middle panel shows the field of view in the IRIS SJI 2796 \AA\ image. This image is just to display the network and inter-network regions in the solar chromosphere above which we have chosen solar TR in Si IV line. The right panel shows the LOS magnetogram.The black vertical line is the position of slit. The region between blue asterisks is the bright network region. The region between green asterisks is the dark inter-network region. Above these chosen regions, we derive TR spectra using the observations of SI IV 1393.755 \AA\ line. }
    \label{fig:figure1}
\end{figure*}

\cite{2009ApJ...696.1448W} studied the Doppler shift and intensity oscillations in five coronal lines and one transition line. They have investigated the phase difference between Doppler shift and intensity oscillations along with temperature dependence of amplitude in both Doppler shift and intensity oscillations. On the basis of in-phase relationship between Doppler shift and intensity, they have suggested the presence of upward propagating slow magnetoacoustic waves in both TR and corona. \cite{2010ApJ...713..573M} also estimated the phase difference between intensity and velocity signals. They have reported both standing and upward propagating slow magnetoacoustic waves in the active region. \cite{2002SoPh..209..265S} studied a coronal spectral line and reported the signature of propagating waves on the basis of the phase relationship. In this work, we evaluate the period of oscillations in the TR spectral line as observed in the QS. On the basis of the phase relationship between the intensity and the Doppler velocity oscillations, we interpret the type of wave modes in the two different regions of QS (i.e., bright and dark regions). Phase relationships between different parameters (intensity (I) and velocity (V) are estimated in order to interpret the nature of oscillations as distinct wave modes. We report observations obtained with the IRIS in the Si IV line. The observations and data analysis are presented in Section \ref{sec:data}. The results are explained in Section \ref{sec:maximum_oscillatory_powers_TR}, and the discussion and conclusions are presented in Section \ref{sec:discussion}.

\begin{figure*}
    
    \includegraphics[trim = 3.5cm 0.0cm 3.5cm 1.0cm,scale=0.9]{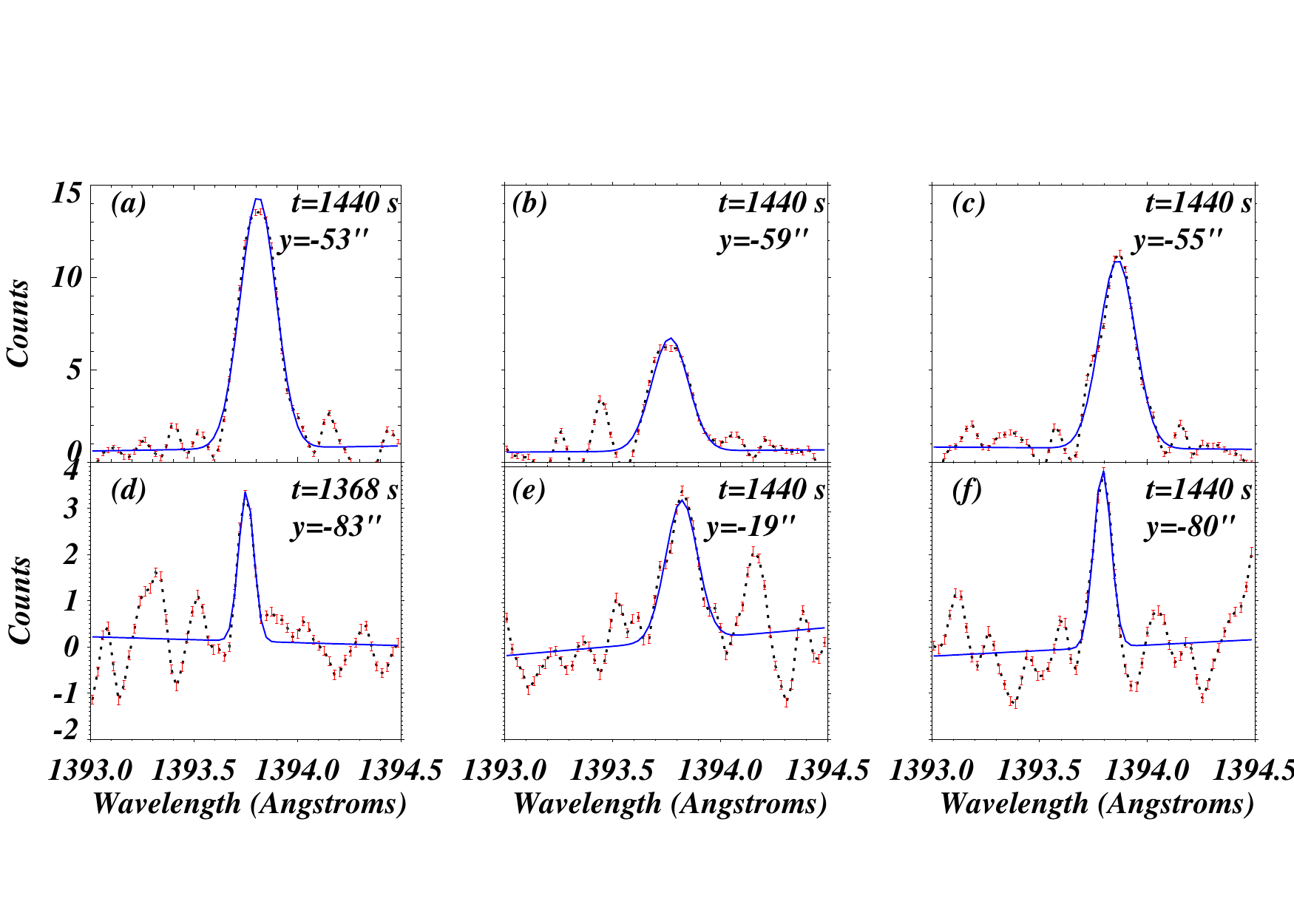}
    \caption{The spectral line profiles of Si IV 1393.755 \AA\ at various locations (dotted black color). The line profiles in the top panel (a, b, and c) belong to the bright TR locations, while the line profiles in the bottom panel (d, e, and f) are derived from dark TR locations. The fitted profiles are obtained using single Gaussian fitting (blue), and the error bars are shown in red.}
    \label{fig:figure2}
\end{figure*}

\section{Observations and data analysis}
\label{sec:data}

We use observations recorded by Interface Region Imaging Spectrograph \citep[IRIS;][]{2014SoPh..289.2733D,2021SoPh..296...84D}. IRIS is a space borne satellite, which provides the spectroscopic data as well as slit-jaw images. Its observations can be made in different spectral lines covering from the solar photosphere, chromosphere to the TR. IRIS observes spectra in three passbands, one in the near-ultraviolet band (NUV) in wavelength range 2783-2834 \AA\, and the other two in the far-ultraviolet bands (FUV1; 1332-1358 \AA\, FUV2; 1389-1407 \AA\ ). Slit-jaw images (SJI) are observed with filters centred on Mg II k 2796 \AA\, the far Mg II h wing at 2832 \AA\ , C II 1330 \AA\, and Si IV 1400 \AA.
In Figure \ref{fig:figure1}, we have also used images of Atmospheric Imaging Assembly \citep[AIA;][]{2012SoPh..275...17L} and Heliospheric Magnetic Imager \citep[HMI;][]{2012SoPh..275..207S} onboard Solar Dynamics Observatory (SDO). AIA observes Sun's atmosphere in UV and EUV with eight different passbands sensitive to plasma at different temperatures. It has a spatial resolution of 0.6" pixel$^{-1}$ in all the wavelength bands (UV and EUV). Different wavebands provide signature of various dynamics happening in various layers of the solar atmosphere. The HMI instrument is also a part of the SDO mission. HMI Dopplergram provides velocity map of plasma motion on the solar surface while HMI magnetogram serves us magnetic map of the photosphere with spatial resolution being 0.5" pixel$^{-1}$.

\hfill \break
We used the IRIS spectral data observed on August 15, 2019 from 21:30 to 22:49 UT using in the sit and stare mode with a 0.33" x 174" slit located near the disk center (at (-1",2"); see Fig. \ref{fig:figure1}) and a cadence of 9 seconds.
Figure \ref{fig:figure1} shows the AIA 171 full disc context image (left-panel) and a zoomed view of the observed region (middle-panel) which has been displayed in the SJI 2796 \AA\ emissions. On the right panel, Line of Sight (LOS) SDO/HMI magnetogram is shown. On the middle panel, the black line represents the slit position, which records the spectra including several emission lines. The chosen bright region lies above a region between marked blue asterisk. The chosen dark region in TR lies above the selected region between green marked asterisk. It is evident from the LOS magnetogram (middle panel) that the bright region corresponds to an enhanced magnetic field, while the dark regions correspond to a weak magnetic field.
We have used two dark regions; one is above the bright region and the other is below the bright region. We used the Si IV observation data, having spatial resolution along the slit (y-direction) as 0.16 $arcsec$. The slit width is 0.33 $arcsec$ across the field of view. The spectral resolution of IRIS is 26 m\AA\ for far-ultraviolet lines and 53 m\AA\ for near-ultraviolet lines \citep{2014SoPh..289.2733D, 2021SoPh..296...84D}. We have studied the nature of oscillations found in the TR using the Si IV 1393.755 \AA\ line formed at temperature log T/K = 4.9. The Si IV is one of the strongest lines formed in the TR. It has a Gaussian shape line profile and it is optically thin.
The observed QS region comprises both network and inter-network regions shown in Figure \ref{fig:figure1} (middle-panel). The bright patches in TR are the region above the chromospheric network regions where the magnetic field is strong. The dark patches in TR are the region above the inter-network regions where the magnetic field is weak. To study the oscillations above the chromospheric network area in the solar TR, we have selected the bright patch between y$\approx$-59" and y$\approx$-37". We have applied average binning of $4\times 2$ (four in time and two in Y) in the selected patch to improve the derived Si IV 1393.755 \AA\ line profiles. The oscillations above the inter-network area are studied by selecting two dark patches. One patch lies from y$\approx$-86" to y$\approx$-61", while the other patch lies from y$\approx$-19" to y$\approx$10". In the dark region, the obtained line profile is weak compared to the line profile obtained in the bright region. To increase the signal, we have applied binning of $4\times 4$ (four in time and four in Y).
\begin{figure*}
	\mbox{
    \centering
    \includegraphics[trim = 3.5cm 0.0cm 3.5cm 1.0cm,scale=0.9]{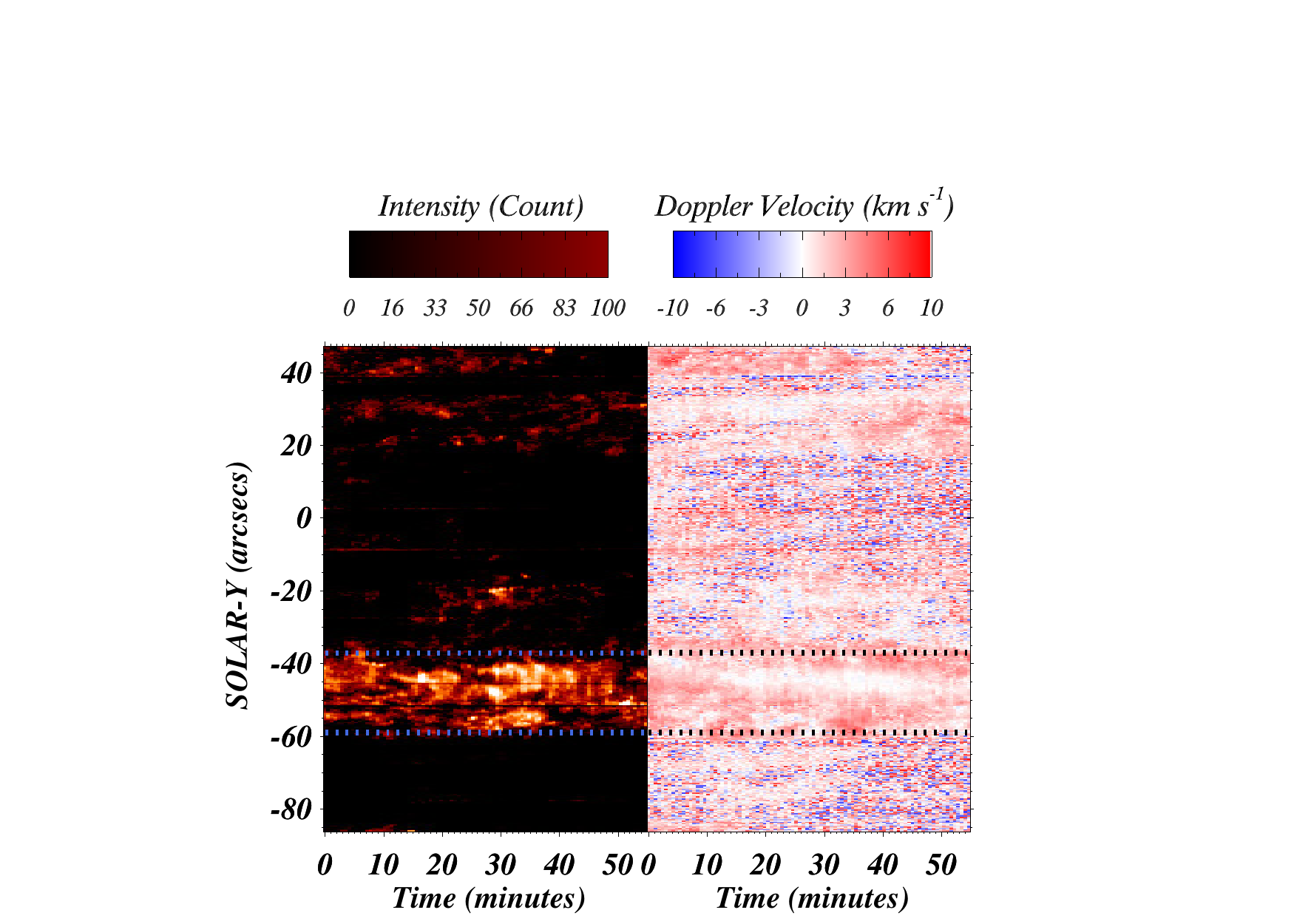}
    \includegraphics[trim = 3.5cm 0.0cm 3.5cm 1.0cm,scale=0.9]{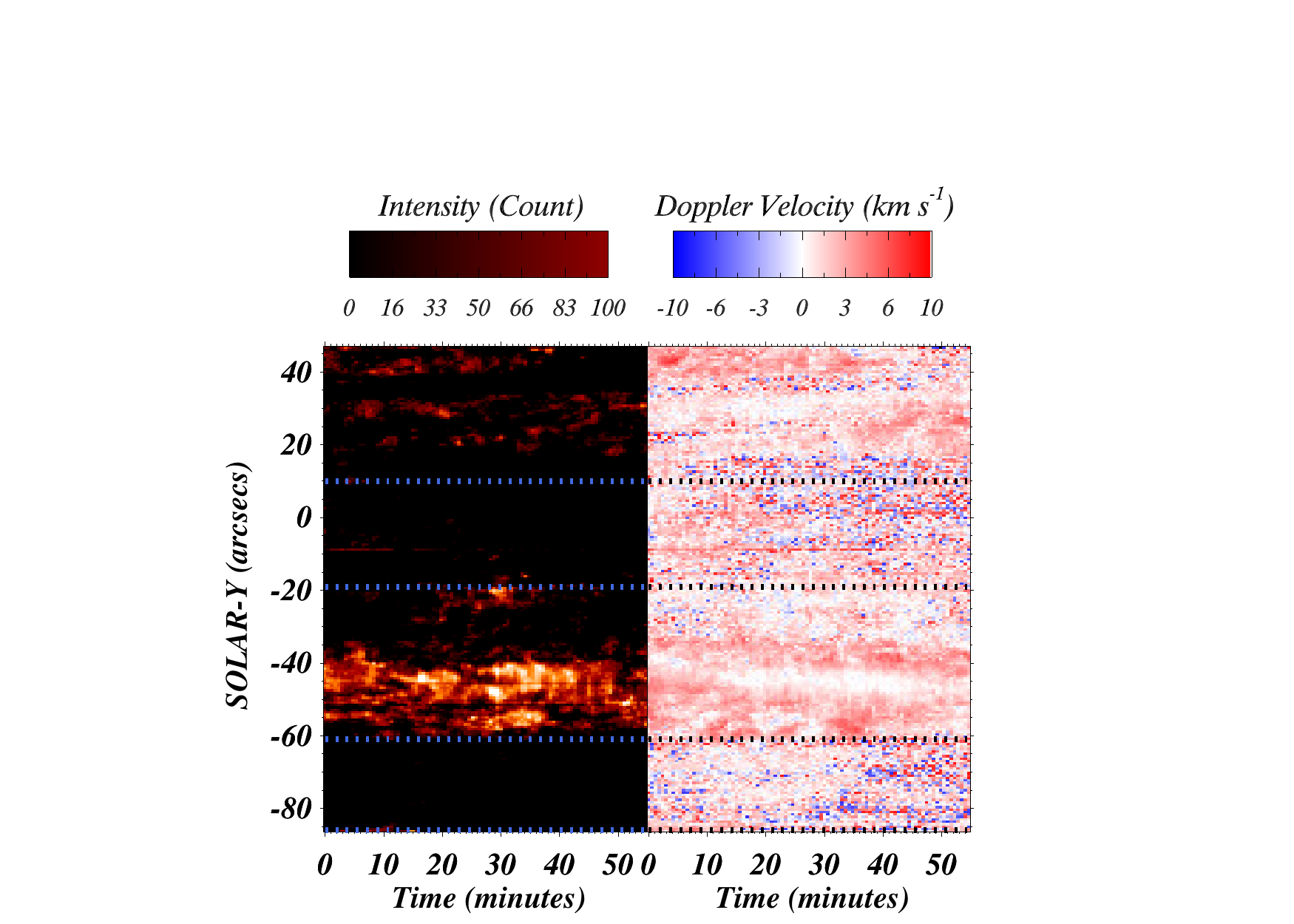}
    }	
	
    \caption{Time-distance map of the intensity and Doppler velocity in Si IV 1393 \AA\ line. The y-axis is parallel to the slit of the instrument.}
    \label{fig:figure3}
\end{figure*}

It is found that the binning processing has clearly enhanced the signal to noise ratio of the observed spectra in the bright regions. We have applied a single Gaussian fitting to extract the spectral line parameters, i.e. intensity, Doppler velocity and line width at each spatial location at various times. We have flagged the missing data of intensity and velocity, and replaced them with the interpolated values. In this way, we constructed the intensity and velocity time-series at each location. However, in the dark TR region, the signal to noise ratio is not improved evidently due to large noises in the data. We have manually checked each spectral line profile at different time steps at each location in this region to verify the reliability of obtained oscillations inherent in the time-series. We have flagged the spectra where signal is absent, instead we interpolated the intensity and velocity for these spectra from the fitted values of nearby locations and times. In the appendix-\ref{spectra}, we have explained the way we have used to derive the spectral parameters. Thereafter, we constructed the time-series of intensity and velocity in dark region.

Figure \ref{fig:figure2} shows some examples of Si IV line profiles with the corresponding Gaussian fittings for the bright region (Panels (a)-(c)), and for the dark region (Panels (d)-(f)). Spectral line profiles are shown in dotted black color, and fitted profiles are overlaid in blue color. The error bars are shown in the red color. To determine the absolute Doppler velocity in Si IV, we first estimate a correction for the given rest wavelength using a cool line nearby as a reference{\footnote{\url{https://pyoung.org/quick_guides/iris_auto_fit.html}}}. We have taken a cool S I 1392 line, because it has very small Doppler velocities (typically < 1 km~s$^{ - 1}$). We have taken the average of the data along the slit at first time step and estimate the centroid of the averaged line profile relative to its rest wavelength. We then add this centroid value (as a correction) to the given rest wavelength of Si IV 1393 and finally obtain the corrected reference wavelength used for the Doppler velocity measurement.

\hfill \break
We used the fitted parameters and generated the maps of the peak intensity and Doppler velocity as shown in Figure \ref{fig:figure3}. We derived time series of intensity and velocity at each 'y' location within the region of interest i.e. chosen bright and dark regions in TR. We performed the power spectral analysis of each time-series and estimate the significant periods at different locations in the region of interest. The details of the scientific results and associated analysis are described in Section \ref{sec:maximum_oscillatory_powers_TR}.


\section{Results}

\label{sec:maximum_oscillatory_powers_TR}

\subsection{Wavelet Analysis and Noise Model Fit}
2
The properties of oscillations in the solar TR can be studied using the wavelet transform. The wavelet transform decomposes time-varying signals into the time-frequency space components, allowing the significant oscillation modes in a power spectrum to be determined. We have performed the wavelet transform on each time-series derived from the selected bright and dark regions in TR.  \cite{1998BAMS...79...61T} provided a detailed description of the wavelet analysis tool. Wavelet transform is the convolution of the time-series with a mother function. We have used Morlet wavelet as a mother function.

\begin{figure*}
  	\mbox{
   \centering
   	\includegraphics[width=.95\linewidth]{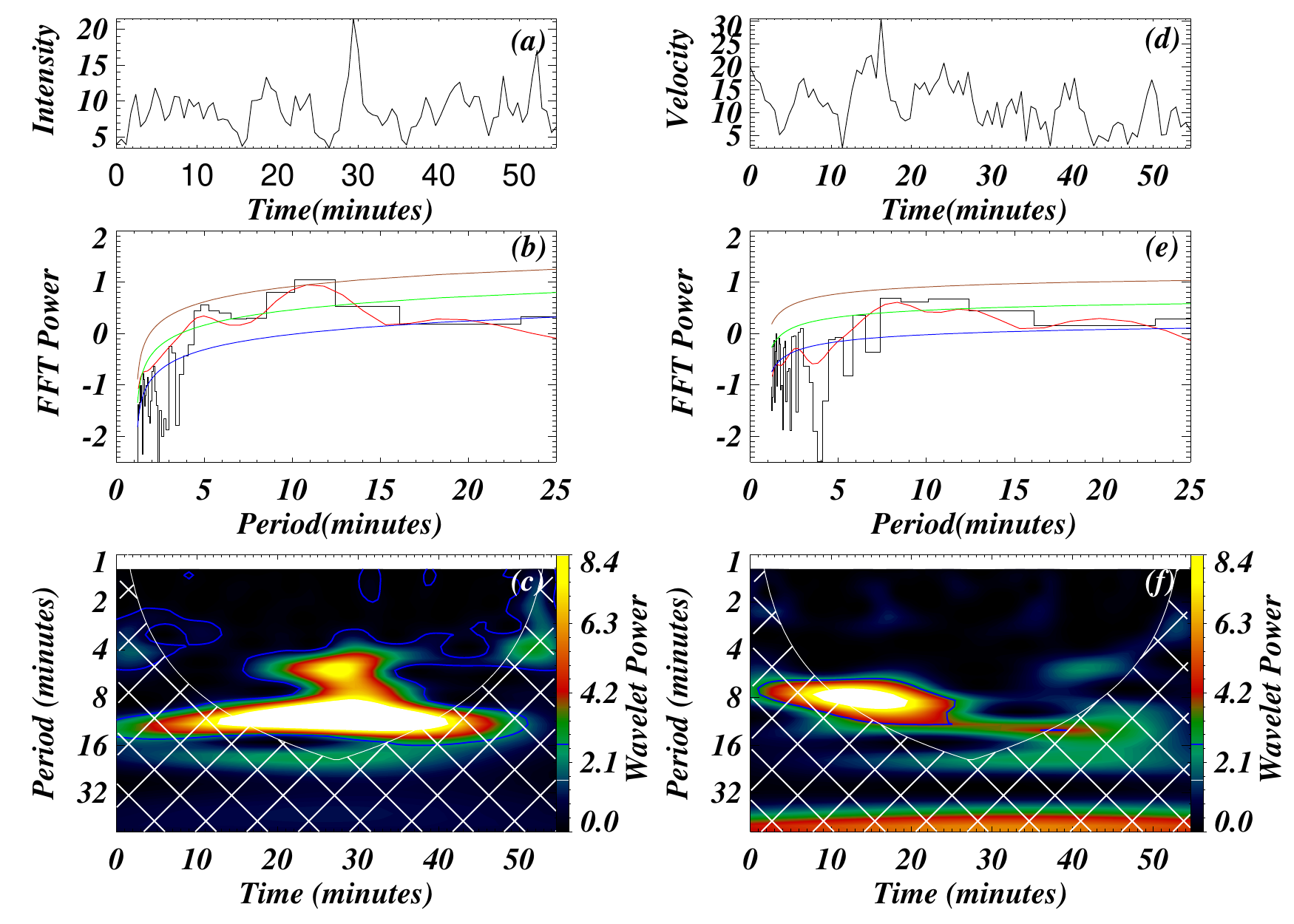} 
   	}

 \caption{The time-series of intensity and Doppler velocity at a bright location (y$\approx$-49") are shown in panels (a) and (d). The model fits for intensity and Doppler velocity are shown in panels (b) and (e). The black curve represents FFT power, while the red curve represents the time averaged global power of the wavelet spectrum. The blue curve depicts the fitted model, which is evaluated using the power law model. Using the fitted model, the 95\% local significance level (green-curve) and 95\% global significance level (brown-curve) are estimated. The wavelet power spectrum of intensity and velocity are shown in panels (c) and (f), respectively. The blue-line contour represents the 95\% global significance level, and the power enclosed within the contour is considered significant.}

  \label{fig:figure4}
\end{figure*}

\begin{figure*}
  	\mbox{
   \centering
   	\includegraphics[width=.95\linewidth]{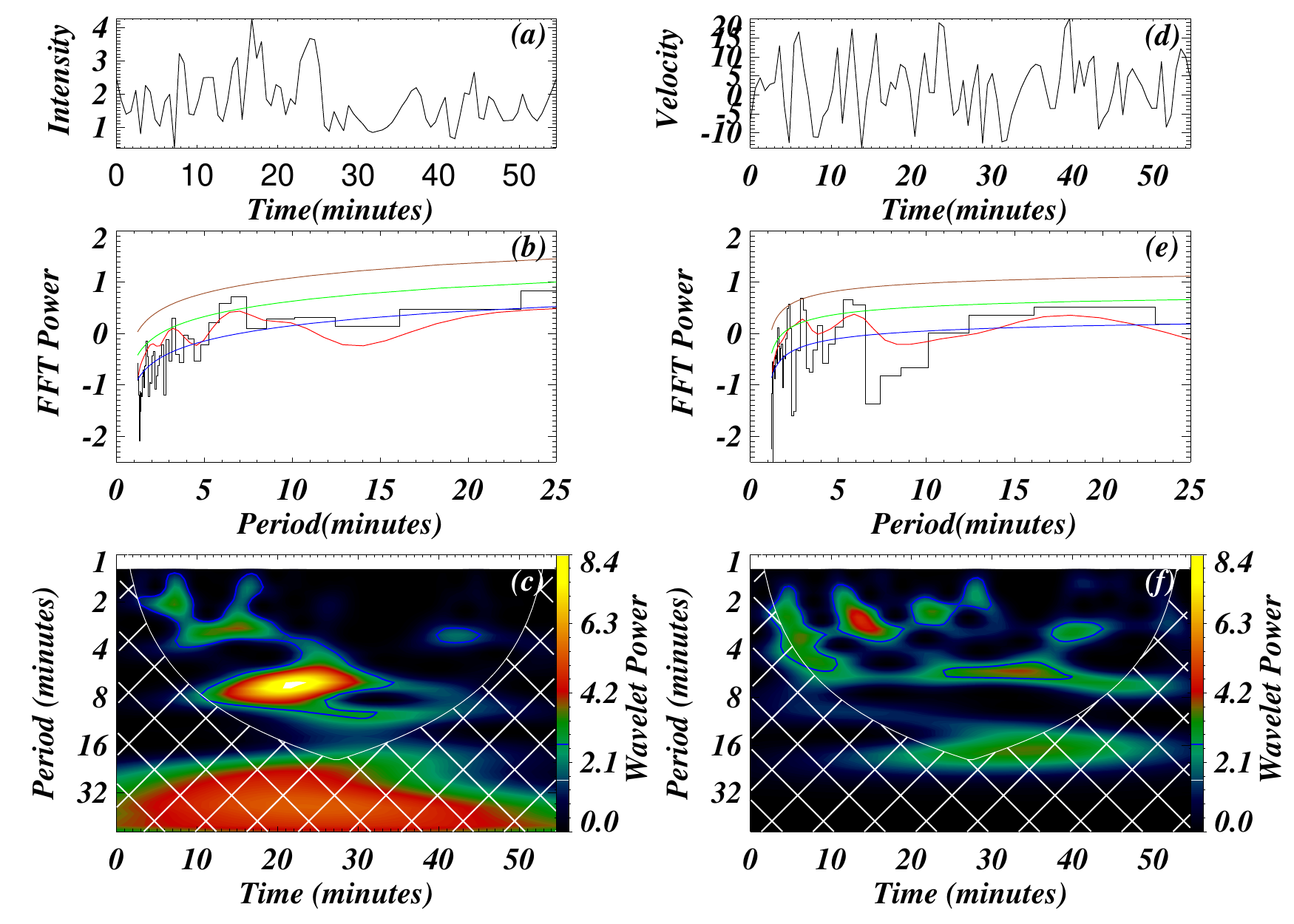} 
   	}
  \caption{ The left panel shows the time-series, model fit, and  power spectrum of  intensity at a chosen dark location (y$\approx$-77"). The right panel shows the time series, model fit, and  the power spectrum of the Doppler velocity time-series. The other description of panels (a)-(f) are same as given in Figure \ref{fig:figure4}.}
  \label{fig:figure5}
\end{figure*}

\hfill \break
The power of a wavelet transform is defined as the square of its absolute magnitude. The global wavelet power is the time-averaged wavelet power, and it may be used to analyze dominating oscillations. Figure \ref{fig:figure4} (c) and \ref{fig:figure4} (f) show the example of the wavelet power spectrum of the intensity time-series and velocity time-series respectively in a bright region at y$\approx$-49", while Figure \ref{fig:figure5} (c) and \ref{fig:figure5} (f) show the example of wavelet power spectrum of the intensity time-series and velocity time-series respectively in a dark region at y$\approx$-77". 
In the wavelet power spectrum, we have set the contour levels with minimum value 0.0 and maximum value 8.45 and the same is shown in the vertical colorbar next to the power spectrum (Figure \ref{fig:figure4} (c), \ref{fig:figure4} (f), \ref{fig:figure5} (c), \ref{fig:figure5} (f)). We have fixed the maximum value of contour level as 8.85, and color corresponding to this power is yellow. Therefore, the power in white color is considered to be greater than the maximum value of the contour level (i.e. 8.85). From Figure \ref{fig:figure4} (c) and \ref{fig:figure5} (c), it is evident that intensity wavelet power at bright location (Figure \ref{fig:figure4} (c)) is greater than the intensity wavelet power at a dark location (Figure \ref{fig:figure5} (c)). We are expecting the higher power at bright locations as signal is high (high signal to noise ration (SNR)) at bright locations as compared to a dark locations which has low SNR.

The cross hatched region shown in the power wavelet is termed as the "cone of influence" (COI), where edge effects become prominent and power inside this region is not reliable \citep{1998BAMS...79...61T}. To improve the reliability of our results, we excluded the power lies in the COI region from our analysis. The maximum allowed period from COI for the given time-series in intensity and Doppler velocity is 19.723 min. Detrending can be used to remove the long period component (greater than the COI) from the time-series. However, as demonstrated by \cite{2016ApJ...825..110A}, detrending may introduce spurious periodicities. As a result, we did not use detrending before applying the wavelet. We subtracted the mean from the time-series, normalised it to its standard deviation, and then computed the wavelet transformation using the \textbf{wavelet.pro} routine. The wavelet power is then calculated by taking the absolute square of the wavelet transformation.

To ensure the reliability of the obtained power, the significance level contour must be produced. If the power is within the contour level, it is significant, otherwise it is not considered as a significant one. The white noise and red noise models have been used to determine significant level by \cite{1998BAMS...79...61T}. However, these models can occasionally provide inaccurate information regarding the significance of the observed power \citep{2016ApJ...825..110A,2020A&A...634A..63K}. To calculate the significant level, therefore we used the method described in \cite{2016ApJ...825..110A} and \cite{2020A&A...634A..63K}. A power law noise equation was modelled by \cite{2016ApJ...825..110A} and can be utilised as a background noise model. The power law noise equation comprises three terms, namely the first is a power law function, the second is a kappa function that is used for the high-energy events, and the third component is a constant. Since we are studying the QS, which is less energetic and free from transients than the event analyzed by \cite{2016ApJ...825..110A}, therefore we did not employ the kappa function. The power law model has also been employed by \cite{2017SoPh..292..165T} and \cite{2020A&A...634A..63K} in their respective analysis. Power law noise function is modeled as 
  
  \begin{equation} \label{eq1}
   \sigma(\nu) = A\nu^s + C 
 \end{equation}

  \hfill \break
  The fitted noise model was created by fitting each Fast Fourier Transform (FFT) power spectrum with the power law function using the \textbf{mpfitfun.pro} function. The significance level is calculated using this noise model. \cite{2016ApJ...825..110A} has explained how to determine the significance level using background noise model. Accordingly, we fitted $\sigma$($\nu$) to FFT of intensity time-series as well as FFT of Doppler velocity time-series at all bright and dark locations in the TR. Figure \ref{fig:figure4} (b) represents an example of intensity power, and Figure \ref{fig:figure4} (e) represents an example of velocity power at a bright location (y$\approx$-49"). Figure \ref{fig:figure5} (b) represents an example of the intensity power, while Figure \ref{fig:figure5} (e) represents an example of the velocity power at a dark location (y$\approx$-77"). The power spectrum of the FFT (black histogram), the time averaged wavelet spectrum (red-line), and the fitted power law noise model (blue-line) are shown in each panel. We calculated the local 95 \% (green-line in both the Figures) and global 95 \%  Fourier confidence levels (brown-line in both the Figures) using this model. The global 95\% wavelet confidence levels are higher than the local 95\% wavelet confidence levels in this case. \cite{2016ApJ...825..110A} provided the methods for determining the significance level using a fitted noise model. We have adopted it and the wavelet power map is outlined by a blue line contour with a 95 \% global significance level determined using the power law model. Significant power lies within a period range, as seen by the wavelet panel (both in the example of bright and dark locations).

\begin{figure*}
  	\mbox{
   \centering
   	\includegraphics[width=.95\linewidth]{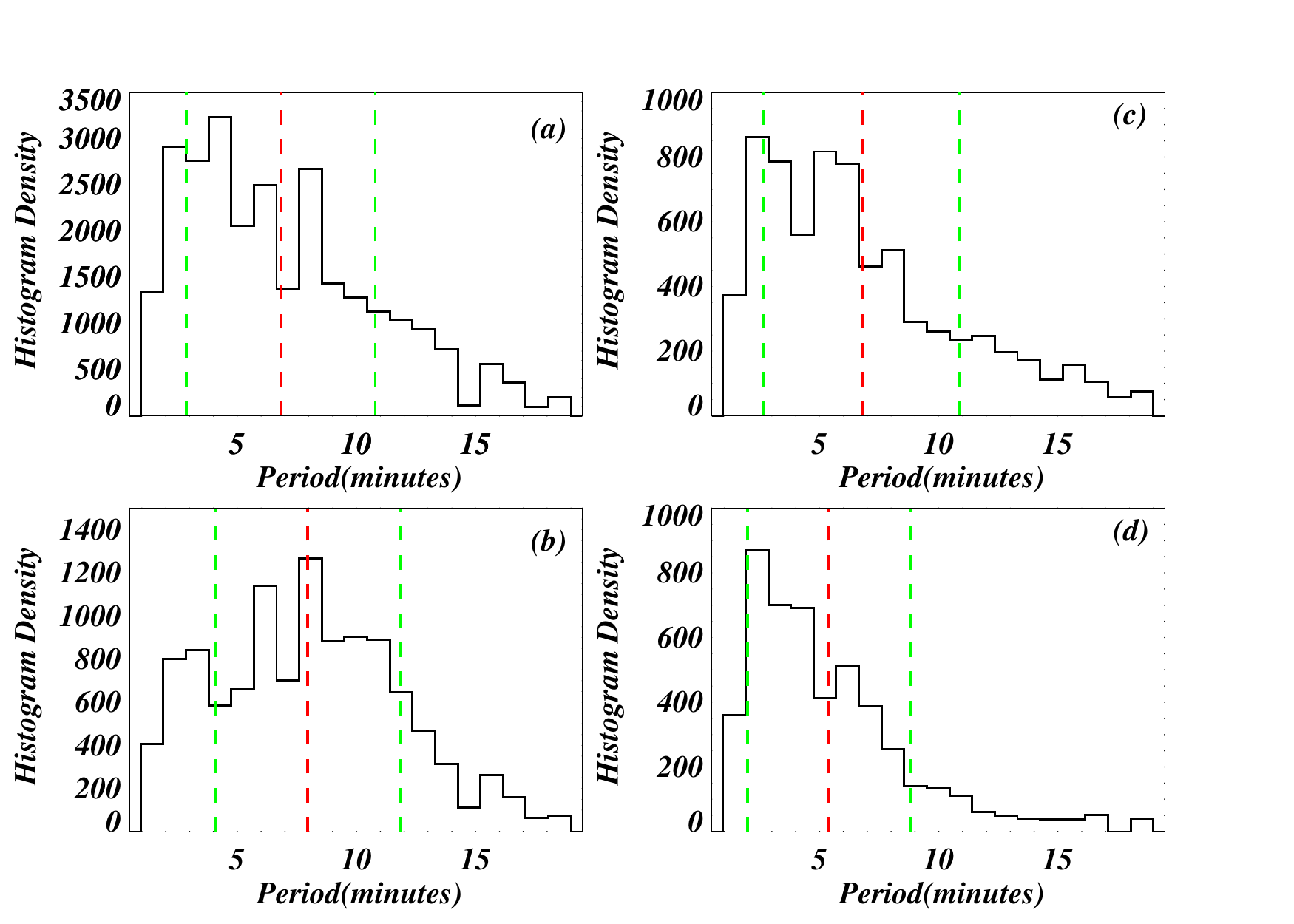} 
   	}
  \caption{The distribution of period in intensity (a) and velocity (b) at bright locations is shown in the left column. The distribution of the period in intensity (c) and velocity (d) at the dark locations is shown in the right column. The red line represents the mean period, and the green line represents the 1-$\sigma$ value.}
  \label{fig:figure6}
\end{figure*}

\hfill \break   
In the bright and dark regions, we examined the intensity and velocity oscillations. The selected bright region is enclosed by the two blue dashed lines in Figure \ref{fig:figure3} (left-panel). We examined all the locations between y$\approx$-59" to y$\approx$-37" and applied the wavelet transform to all the derived time-series of intensity and velocity. Figure \ref{fig:figure4} shows a representative case of wavelet analysis of a time-series (intensity and Doppler velocity) derived from a position at y$\approx$-49". Figures \ref{fig:figure4} (a) and \ref{fig:figure4} (c) show the intensity time-series and the associated wavelet power spectrum, respectively. Figures \ref{fig:figure4} (d) and \ref{fig:figure4} (f) show the velocity time-series and the associated wavelet power spectrum, respectively. We constructed the fitted noise model by fitting the FFT of intensity time-series and Doppler velocity time-series with the power law function as stated above. To obtain the fitted noise model at each time-series, we have fitted all the intensity and velocity power in the bright region. We removed some specific locations from the analysis since the model did not fit the FFT signal well in those locations. The significant power is found in the range 3-16 of min in the intensity time-series, and 5-10 of min in the velocity time-series, as shown by the power wavelet. To extract the reliable period of the intensity and velocity signals, we utilised the following conditions:
(i) Only significant power is taken (i.e., wavelet power lies within the regime of 95\% global confidence level); (ii) To avoid any edge effect, a region outside the COI is taken. As a result, we have collected all the period values that satisfy the aforementioned criteria. We have applied the same conditions to all of the intensity and velocity wavelet spectra and extracted all the reliable and significant periods. We presented a 1-D histogram (i.e., period vs frequency) of all the bright locations after estimating the period information (see, for example, Figure \ref{fig:figure6}). Figure \ref{fig:figure6} (a) shows the distribution of period in the intensity, and Figure \ref{fig:figure6} (b) depicts the distribution of period in the velocity. We have then calculated the mean period of each case with a 1-$\sigma$ variation. Mean period in intensity oscillations is 7.0 $\pm$ 3.9 min, and the mean period in velocity is 8.0 $\pm$ 3.9 min. In the bright region, intensity period is consistent within velocity period with a 1-$\sigma$ range. The mean period is shown in green colour and the 1-$\sigma$ is shown in the red colour in Figure \ref{fig:figure6}.


\hfill \break
Now, we analyse the oscillations present in the intensity and velocity time-series derived from the dark region. In the right panel of Figure \ref{fig:figure3}, we show the regions that we have selected for the study of oscillations in the dark TR region. We have chosen two patches for the analysis. One region lies between y$\approx$-86" and y$\approx$-61", and the other ranges from y$\approx$-19" to y$\approx$10". The data in the dark region was noisy and we have manually checked each spectra to make sure whether the obtained oscillations are reliable or not. The details of the analysis is described in the Appendix-\ref{spectra}. In total there are 80 locations along the slit in the selected dark regions, and out of which at around 35 locations the signal was poor. Therefore, we could not obtained the time series from those locations so we have excluded them from the analysis. We have performed a wavelet analysis on the obtained time-series from the chosen locations. A representative case of the wavelet analysis of a time-series is shown in Figure \ref{fig:figure5}, which is located at y$\approx$-77". Figure \ref{fig:figure5} (a) shows the intensity time-series and Figure \ref{fig:figure5} (c) shows the corresponding wavelet power spectrum. Similarly, Figure \ref{fig:figure5} (d) shows the velocity time-series and Figure \ref{fig:figure5} (f) shows the corresponding wavelet power spectrum. Both the intensity and velocity oscillations correspond to the same location. We have used the same methodology as described above to calculate the fitted power law model and 95\% local and global confidence level. We have omitted those locations where power law model does not fit to the FFT power, like the way we have done in the bright region. The wavelet power lies within 95\% global confidence contour (blue-line contour) is the significant power in the given location. We have applied the same conditions here as mentioned above for the bright region to all the locations in intensity and velocity and extracted the period information. Figure \ref{fig:figure6} (c) shows the distribution of period in intensity and Figure \ref{fig:figure6} (d) shows the distribution of period in velocity. Mean period in intensity is 7.0 $\pm$ 4.1 min. The mean period in velocity is 5.4 $\pm$ 3.4 min. Here, distribution is also consistent within the 1-$\sigma$ limit as we have seen is the bright region. The statistically significant periods in the velocity and intensity of multiple bright and dark locations in the solar TR most likely imply the presence of MHD oscillations. To further investigate the physical nature of these oscillations, in the forthcoming sub-section \ref{sec:phase_analysis_TR}, we perform the phase-analyses between intensity and velocity.

\subsection{Cross Power between Intensity and Velocity time-series}
\label{sec:cross_power_TR}

The cross-wavelet spectrum of two time-series may also be computed, which provides information on their correlations and phase differences \citep[e.g.,][]{1998BAMS...79...61T,2004ApJ...604..936B,2004ApJ...604..924M,2017ApJS..229...10J,2018MNRAS.479.5512K}. The cross-spectrum reveals high-common-power and time-frequency spaces present in the two time-series, whereas the wavelet coherence identifies regions where the time-series are coherent but not necessarily have a high common power. Wavelet coherence is required to detect the co-movements of the oscillations present in two time-series. Wavelet coherence has a minimum value of zero, which shows that there is no correlation between two time-series. Wavelet coherence has a maximum value of one, indicating that there is the highest correlation between two time-series. For the present analysis, the phase difference between intensity and velocity time-series provide information about the physical nature of oscillations present in TR at different chosen locations. The complex and the real arguments of the cross-spectrum are used to estimate the phase-lags. The cross-wavelet, wavelet coherence, and phase difference between these two time-series are evaluated using \textbf{wave coherency.pro} module in SSW IDL. We estimated the cross wavelet which is the multiple product of intensity wavelet and the complex conjugate of velocity wavelet. We also estimated cross wavelet power which is the square of the absolute magnitude of the cross wavelet array between intensity and velocity at all locations in bright and dark regions \citep{1998BAMS...79...61T,2004ApJ...604..936B}. Coherence is defined as the normalisation of the cross wavelet power by multiplying the powers of both time-series \citep{1998BAMS...79...61T,2004ApJ...604..936B}. The imaginary and real components of cross-wavelet power can be used to compute phase differences \citep{1998BAMS...79...61T,2004ApJ...604..936B}. We computed the cross FFT, defined as multiplication of FFT of one time-series with the complex conjugate of FFT of the other time-series. From the cross FFT, we evaluated the fitted noise model and 95\% global significance level.

\begin{figure*}
  	\mbox{
   \centering
   	\includegraphics[width=.95\linewidth]{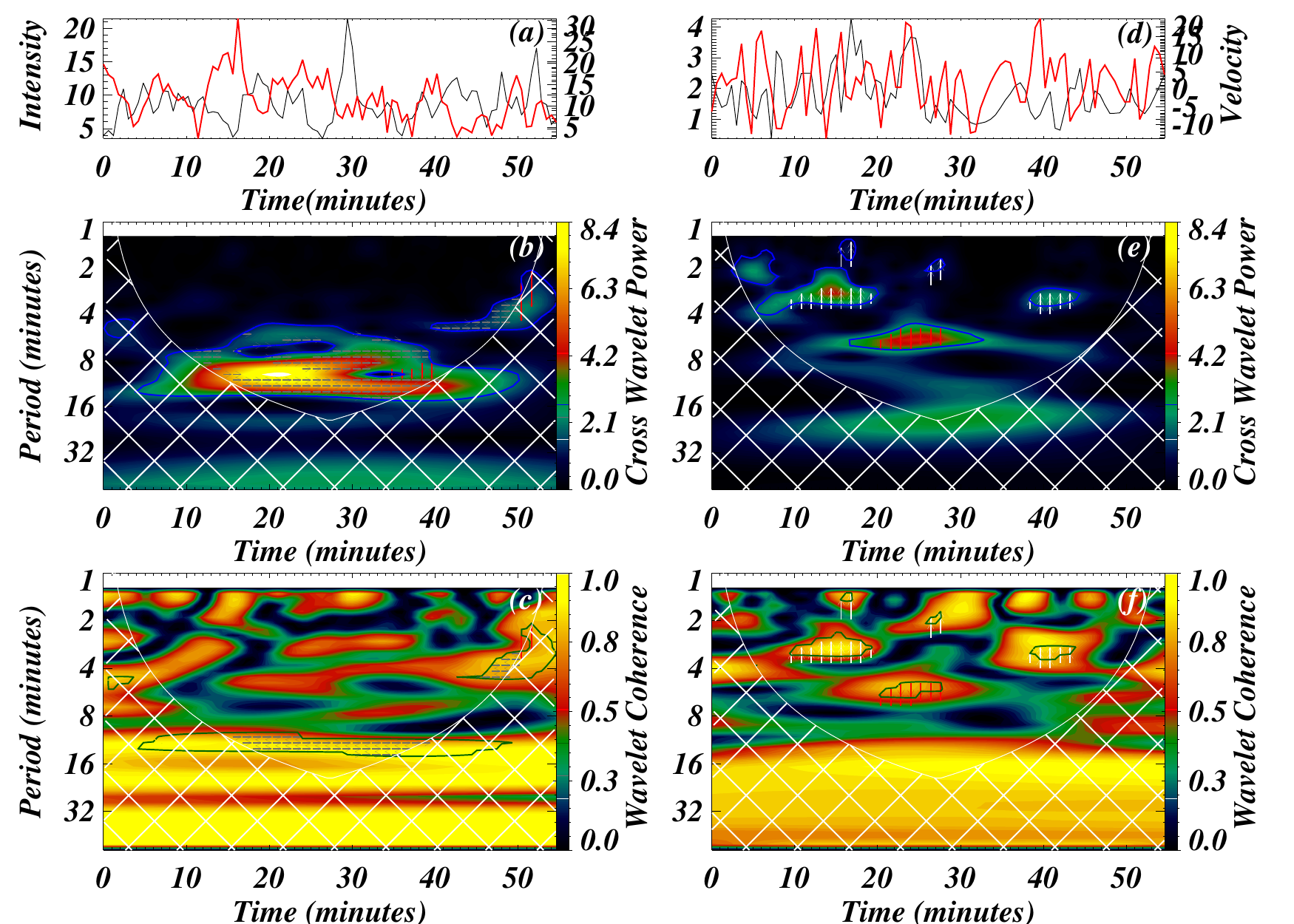} 
   	}
 \caption{The time-series of intensity (black) and Doppler velocity (red) in the representative bright and dark regions are shown in panels (a) and (d), respectively. The cross power wavelet between two time series is shown in panels (b) and (e) for the bright and dark regions, respectively. The blue line contour represents the global significance level of 95 \%. The phase differences are over plotted within the significant contour. The grey arrows represent the standing waves, red arrows represent the downward propagating waves and white arrows represent the upward propagating waves. The wavelet coherence between two time-series is shown in panels (c) and (f). The green line contour represents a global significance level of 95\% with wavelet coherence greater than 0.70. The arrows of phase differences are overlaid in the high coherence regime (i.e., coherence above 0.70).}
  \label{fig:figure7}
\end{figure*}

We have shown one representative case of the bright region in Figure \ref{fig:figure7} (left-panel) and one representative case of the dark region in Figure \ref{fig:figure7} (right-panel). Figure \ref{fig:figure7}
 (a) shows the intensity time-series in black color and velocity time-series in red color. Figure \ref{fig:figure7} (b) shows the cross wavelet power, and Figure \ref{fig:figure7} (c) the wavelet coherence between intensity and velocity time-series of a chosen bright location (y$\approx$-49"). Similarly, Figure \ref{fig:figure7} (d) shows the intensity and velocity time-series for the dark region. Figure \ref{fig:figure7} (e) shows the cross wavelet power, and Figure \ref{fig:figure7} (f) the wavelet coherence between intensity and velocity time-series of a chosen dark location (y$\approx$-77"). Blue line demonstrates the wavelet power with the 95\% global significance level, and the power lies within the contour is considered significant. Further, we have applied two more conditions i.e., (i) Region outside the COI is considered, (ii) Threshold value of wavelet coherence is taken 0.7 and plotted 95 \% global significance level contour on the wavelet coherence map and the same is shown in the green color.

\subsection{Analysis of phase difference between intensity and velocity time-series and interpretation of wave modes}
\label{sec:phase_analysis_TR}

Phase differences between intensity and velocity signals can be used to understand the oscillatory behaviour of the solar atmosphere. Phase difference between intensity and Doppler velocity signals are important factors in determining the physical nature of the oscillations. According to linear MHD wave theory, density (hence intensity) and velocity oscillations with periods of some minutes are usually associated with slow magnetoacoustic longitudinal waves (see, Table \ref{table:phase}). The objective of this analysis is to identify the MHD modes associated with the detected oscillations in QS. The ideal phase relationships between intensity and velocity can be used to determine whether a slow wave is propagating or standing (see, Table \ref{table:phase}). The propagating waves show an in-phase (or out of phase) variation when viewed along their direction of propagation, whereas standing waves show a quarter-period variation between intensity and velocity \citep[e.g.,][]{2002SoPh..209..265S,2010ApJ...721..744K}. \cite{2009ApJ...696.1448W} has also provided the equation for velocity and density perturbations. The authors have estimated the relation between velocity and density perturbations and suggested that for the upward propagating waves the velocity and density perturbations are in phase, while for downward propagating waves these perturbations are in the opposite phase. The velocity perturbations are related to the Doppler velocity signal while the density perturbations are related to the intensity signal of the observations. They have computed the phase difference between intensity and Doppler shift oscillation in five coronal lines and one transition line and found that the phase of intensity oscillation is slightly earlier (by about 20\degree - 30\degree) than the Doppler shift oscillation. They interpreted these oscillations as the signature of upwardly propagating slow magnetoacoustic waves in both TR and corona near the footpoint of a loop on the basis of approximate phase relations. In the previous studies, standing slow waves are also reported based on phase studies between intensity and velocity. The presence of damped oscillatory Doppler shifts was found in the hot coronal loop observations by Solar Ultraviolet Measurements of Emitted Radiation (SUMER) spectrometer on Solar and Heliospheric Observatory (SOHO). These oscillations were interpreted as standing slow waves on the basis of their phase speed and the phase shift between intensity and velocity oscillations. The phase shift between intensity and velocity oscillations was quarter period \cite[e.g.,][and references therein]{2003A&A...402L..17W,2003A&A...406.1105W,2005A&A...435..753W,2011SSRv..158..397W}.


 \begin{table}
\caption{Ideal I-V Phase in slow magnetoacoustics waves} 
\centering 
\begin{tabular}{c c c } 
\hline\hline 
Mode & Phase Difference & Reference \\ [0.5ex] 
\hline
Propagating wave & 0, $\pm\pi$  & \cite{2009ApJ...696.1448W} \\ 

 &  & \cite{2010ApJ...721..744K} \\[1ex]

\hfill \break
Standing wave & $\pm\pi/2$ & \cite{2008ApJ...681L..41M} \\ 
 

 &  & \cite{2011SSRv..158..397W}  \\[1ex]
\hline 
\end{tabular}
\label{table:phase} 
\end{table}


Theoretical modeling of the magnetoacoustic waves in coronal loops has been done previously in a number of studies to see the variation of phase shift between density and velocity perturbations with respect to the background densities and temperatures. Several damping mechanism play an important role in the propagation or evolution of these magnetoaccoustic oscillations such as thermal conduction, compressive viscosity, and radiative cooling along with heating-cooling imbalance \citep[e.g.,][]{2009A&A...494..339O,2021SoPh..296...20P,2021SoPh..296..105P,2022SoPh..297....5P}. It is well known from theoretical models and simulations that thermal conduction can lead to a phase shift between velocity and density perturbations of magnetoaccoustic oscillations. From the 1D simulation \citep[][Fig. 1]{2019ApJ...886....2W} or 1D analytical models \citep[e.g.,][]{2021SoPh..296...20P,2021SoPh..296..105P,2022SoPh..297....5P} of phase shifts in slow magnetoaccoustic oscillations it is known that thermal conduction causes a phase shift in density (thus intensity) earlier than velocity. 
In order to analyze our observations in the context of the slow magnetoaccoustic waves (both propagating and standing), we  consider a 0.5D coordinate system with the origin at the upper boundary and the positive direction of the axis toward the bottom layer of the solar atmosphere. We assume the Doppler downflow (red shift) to be defined as the positive values. Assuming the linear velocity perturbation to be of sinusoidal nature as below, we can write

 \begin{equation} \label{eq2}
  V = V_0\ sin(\omega t).
 \end{equation}

Using above equation we can write the variation of density perturbation as follows :
\begin{itemize}

\item In standing wave,

 \hfill
 
 For a location near upper end in solar TR,
 \begin{equation} \label{eq3}
  \rho = \rho_0\ sin(\omega t+\pi/2+\phi), 
 \end{equation}
 
 \hfill
 
 , while for a location near lower end in TR,
 \begin{equation} \label{eq4}
   \rho = \rho_0\ sin(\omega t-\pi/2+\phi), 
\end{equation}

\hfill \break 
\item In propagating wave, 
\hfill \break
for the downward propagating wave in the solar TR,
  \begin{equation} \label{eq5}
   \rho = \rho_0\ sin(\omega t+\phi), 
 \end{equation}
 
  \hfill 
  
, while for the upward propagating wave, we can write    
  \begin{equation} \label{eq6}
    \rho = \rho_0\ sin(\omega t+\pi+\phi)\quad or \quad\rho = \rho_0\ sin(\omega t-\pi+\phi), 
 \end{equation}
 
  \hfill

\end{itemize}
  where 0 < $\phi$ < $\pi$/2\ is the phase of $\rho$ relative to the velocity caused by the dissipative effects in the plasma medium.

\begin{figure*}
   	
   	\includegraphics[width=.9\linewidth]{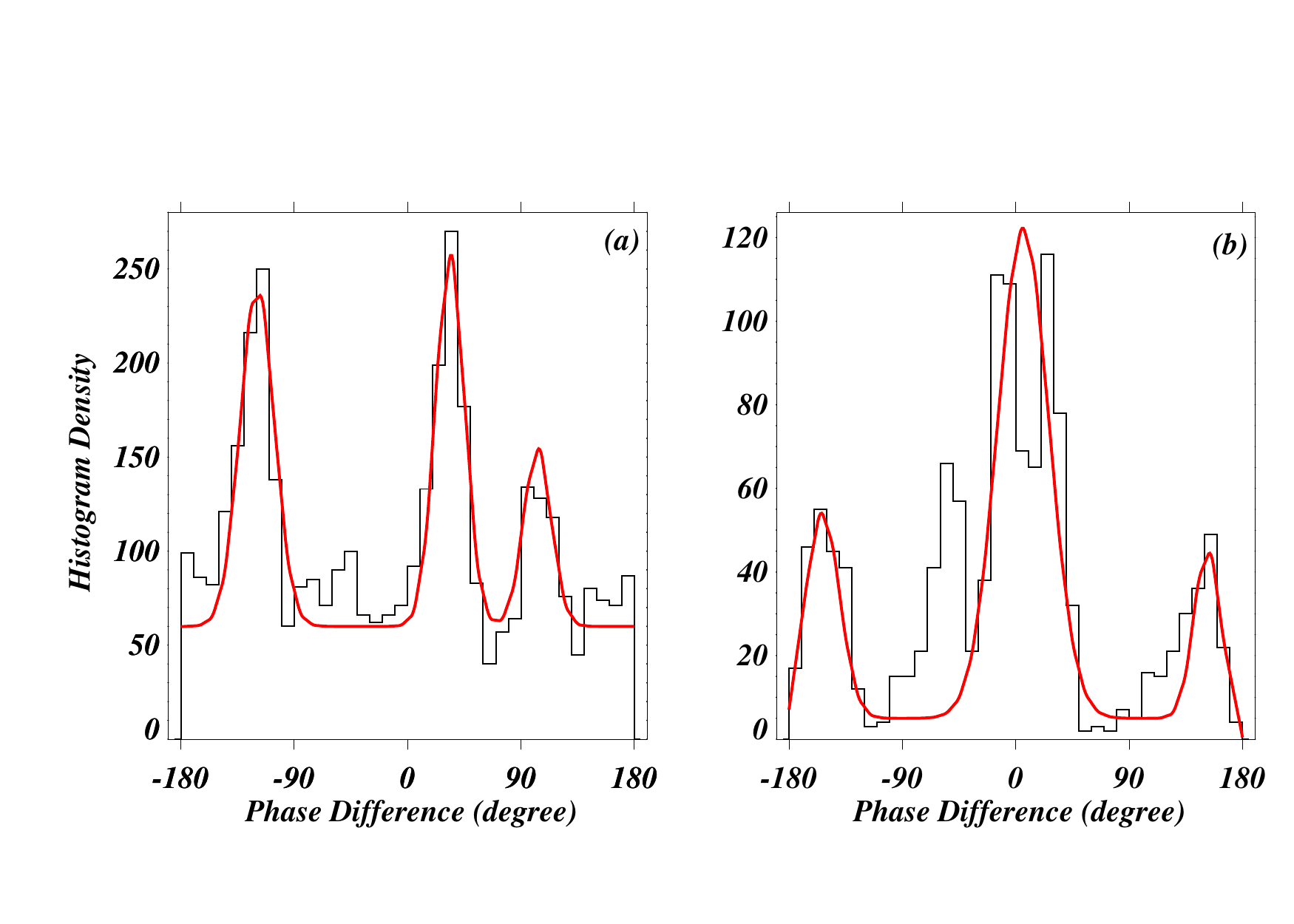}

 \caption{The distribution of phase difference in a bright (a) and dark (b) regions, which ranges between -180\degree and 180\degree. The distribution peaks at three phase difference values in both the cases. The over plotted red line is the fitted Gaussian function, which is used to compute the phase difference values at associated with various peaks.}
  \label{fig:figure8}
\end{figure*}

In the present paper, we have used wavelet phase coherence analysis to examine the phase difference between intensity and Doppler shift oscillations in the solar TR. The phase difference between intensity and Doppler velocity has been calculated at each location. The wavelet phase difference is calculated using the real and imaginary components of the complex cross-wavelet transform \citep{1998BAMS...79...61T,2004ApJ...604..936B}. The range of possible phase difference values is -180\degree\ to 
 +180\degree. We have improvised certain conditions to extract the reliable phase difference value, e.g., (i) only those region of the wavelet are taken where cross power is significant (i.e., wavelet power lies within the regime of 95\% global confidence level), (ii) region outside the COI is taken to prevent any edge effect, (iii) threshold value of wavelet coherence is taken 0.70. We have applied these conditions to each location in the TR and extract all the reliable phase difference between intensity and velocity time-series. We have plotted the phase distribution using all the reliable phase values for each loaction. Figure \ref{fig:figure8} (a) shows the distribution of phase at bright locations, and Figure \ref{fig:figure8} (b) shows the distribution of the phase at dark locations. In both the regions, the distribution varies between -180\degree\ to +180\degree. We have applied Gaussian fitting and computed the phase values at three peaks.
In the bright region, the first peak of the distribution lies at -119\degree $\pm$ 13\degree, the second peak lies at 33\degree $\pm$ 10\degree, and the third peak lies at 102\degree $\pm$ 10\degree. In the dark region, the first peak of the distribution lies at -153\degree $\pm$ 13\degree, the second peak lies at 6\degree $\pm$ 20\degree, and the third peak lies at 151\degree $\pm$ 10\degree. On the basis of this distribution these observed significant oscillations in intensity and velocity at TR can be associated with standing slow waves, upward propagating slow magneto-acoustic waves, or downward propagating magnetoacoustic waves.

We can clearly see that a 0.5D uniform loop model for standing or propagating slow magnetoacoustic waves may lead to a prediction of only two distinct peaks in phase shift.


 \begin{table*}
\caption{Theoretical predicted phase differences and their comparison with observed values} 
\centering 
\begin{tabular}{c c c } 
\hline\hline 
PEAK & BRIGHT REGION & DARK REGION \\ [0.5ex] 
\hline
OBSERVED PHASE PEAKS & -119\degree\ $\pm$ 13\degree  & -153\degree\ $\pm$ 13\degree \\[1ex] 

 & 33\degree\ $\pm$ 10\degree  & 6\degree\ $\pm$ 20\degree \\[1ex]
 
 & 102\degree\ $\pm$ 10\degree  & 151\degree\ $\pm$ 10\degree \\[1ex]

\hfill \break

PEAKS FOR PROPAGATING WAVES & $\phi$ = 33\degree\ (assumed) & $\phi$ = 6\degree\ (assumed) \\ [1ex]
 &  $\phi - 180\degree$ = -147\degree & $\phi - 180\degree $ = -174\degree \\[2ex]

\hfill \break 
 
PEAKS FOR STANDING WAVES & $90\degree+\phi$ = 102\degree\ (assumed) & $90\degree+\phi$ = 151\degree\ (assumed) \\ [1ex]
 & $-90\degree+\phi$ = -68\degree & $-90\degree+\phi $ = -29\degree \\[2ex]

\hline 
\end{tabular}
\label{table:peak} 
\end{table*}

 We have observed three peaks in both bright and dark regions. In the bright region, the first peak of the distribution lies at -119\degree $\pm$ 13\degree, the second peak lies at 33\degree $\pm$ 10\degree and the third peak lies at 102\degree $\pm$ 10\degree. On the basis of the standing wave, from third peak of 102\degree, we have obtained $\phi$ = 102\degree - 90\degree = 22 \degree using Eq. (\ref{eq3}). Further using Eq. (\ref{eq4}), we expect the other peak to be at -90\degree\ + $\phi$ = -68\degree. Similarly, on the basis of propagating wave, for the second peak of 33\degree, we obtained $\phi$ = 33\degree\ for the downward propagating wave using Eq. (\ref{eq5}). Further we expect the phase shift for the upward propagating wave to be at $\phi$-180\degree = -147\degree using Eq. (\ref{eq6}).

In the dark region, the first peak of the distribution lies at -153\degree $\pm$ 13\degree, the second peak lies at 6\degree $\pm$ 20\degree and the third peak lies at 151\degree $\pm$ 10\degree. On the basis of standing wave, from third peak of 151\degree, we have obtained $\phi$ = 151\degree - 90\degree = 61\degree using Eq. (\ref{eq3}). Further from Eq. (\ref{eq4}), we expect other peak to be at -90\degree + $\phi$ = -29\degree. Similarly on the basis of the propagating wave, from third peak of 6\degree, we obtained $\phi$ = 6\degree for downward propagating wave using Eq. (\ref{eq5}), and then we predict the phase for the upward propagating wave to be $\phi$ - 180\degree = -174\degree using Eq. (\ref{eq6}). In Table \ref{table:peak}, we summarize the predictions of various peaks of the phase shift values compared to our observations.

We find that the prediction using propagating slow wave model leads to a phase peak at -147\degree\ for the bright region which is close to the first observed peak in the bright region ranging between -106\degree\ to -132\degree. Similarly in the dark region, the prediction using propagating modes leads to a phase peak at -174\degree\ which is again close to the first observed peak in the dark region ranging between -140\degree\ to -166\degree. However, there is still a significant difference in the predicted and observed phase shifts in both bright and dark regions (see, Table \ref{table:peak}). The predicted phase shifts using standing modes is even further away from the observed values and cannot explain the observed results.
The above analysis suggests that the predictions using the 0.5D model cannot explain the observed three distinct peaks of the phase shifts of our analysis.


In Figure \ref{fig:figure7}, we have drawn arrows indicating phase differences on both the cross power and wave coherence maps. We have defined the ranges for the different wave modes in order to draw phase difference arrows using its statistical distribution. We have taken the range of $\pm$3$\sigma$ error to define different wave modes. For example, in the bright region, the first peak lies at -119\degree and third peak lies at 102\degree, both may corresponding to standing waves as they are close to $\pm$ 90\degree\ (see, Table \ref{table:phase}). So, we have defined standing waves mode as a phase difference in the range from -119\degree - 39\degree\ to -119\degree + 39\degree\ or from 102\degree - 30\degree\ to 102\degree + 30\degree. The second peak lies at 33\degree, and it may correspond to downward propagating waves as it is close to 0\degree. The downward wave mode can be defined with the observed phase difference varies from 33\degree - 30\degree\ to 33\degree + 30\degree\ in the bright region. Similarly, we have defined the phase differences range using the peaks of distribution of phase differences in the dark region (see, Figure \ref{fig:figure8} (b)). In the dark region, the first peak lies at -153\degree, and the third peak lies at 151\degree. Both these observed values may correspond to the upward propagating waves as they are close to $\pm$ 180\degree. Therefore, the upward propagating wave mode can be defined as a phase difference in the range from -153\degree - 39\degree\ to -153\degree + 39\degree\ or from 151\degree - 30\degree\ to 151\degree + 30\degree. The second peak lies at 6\degree, therefore, the downward propagating wave modes can be defined as phase difference in the range from 6\degree - 60\degree\ to 6\degree + 60\degree\ in the dark region. Using the defined ranges of different wave modes considering $\pm$ 3$\sigma$ deviation around the observed peaks, we have drawn the phase difference arrows on the cross power map (see, Figure \ref{fig:figure7} (b) and Figure \ref{fig:figure7} (e)). We have plotted a phase difference arrow only in the area associated with 95 \% global significance level. The downward propagating waves are shown by red color, the upward propagating waves are shown by white color, and the standing waves are shown by the grey color arrows. Further, we have displayed the phase difference arrows on the wavelet coherence map in the high coherence regime outside COI (see, Figure \ref{fig:figure7} (c) and Figure \ref{fig:figure7} (f)). In Figure \ref{fig:figure7} (c) grey and red arrows are plotted. In the Figure, the red arrows represent the downward propagating waves, while the grey arrows represent the standing waves. Different wave modes (downward propagating and standing waves) exist at different times and frequencies. Similarly, phase arrows are shown in the representative example of a dark region. In Figure \ref{fig:figure7} (f) the white and red arrows are plotted. In this diagram, the red arrows represent downward propagating waves and the white arrows represent upward propagating waves. This case also demonstrates that different wave modes (upward propagating and downward waves) exist at different times and frequencies. In the shown example of a bright region, the signature of the downward propagating waves is not prominent, therefore in the Appendix-\ref{phase-arrows}, we have shown other example locations in the bright region where the signature of downward propagating wave is clearly visible. We have also shown some others examples from the dark region in the Appendix-\ref{phase-arrows}.

\section{Discussion and conclusions}
\label{sec:discussion}

We used IRIS data to investigate the physical nature of the observed oscillations in the TR of a QS region, and the spectral line used for the study is SI IV 1393 \AA. We calculated the significant periods of intensity oscillations and Doppler velocity oscillations statistically in two different regions (bright and dark regions). We used wavelet analysis to obtain the distribution of periods in the intensity and velocity oscillations in the bright (above network) and dark (above inter-network) regions in the quiet TR.

\hfill \break
We calculated the mean period of the distribution with 1-$\sigma$ uncertainty. We observed long period oscillations dominating in both the regions (> 5 min). The mean period of intensity oscillations in the bright region is about 7 min, and the mean period of velocity oscillations is about 8 min. The mean period of intensity oscillations in the dark region is about 7 min, and the mean period of velocity oscillations is about 5.4 min. We found that the mean period in both the intensity and velocity oscillations is consistent within the 1-$\sigma$ error in both the regions. We aimed to examine the nature of oscillations in the TR, but not to find the origin of these oscillations. We estimated the presence of various wave modes using the phase relationship between intensity (I) and Doppler velocity (V). These TR oscillations have the characteristics of a slow magneto-acoustic wave (either propagating waves or standing waves). The slow waves are generally compressive in nature and propagate along magnetic field lines. These waves are characterized by periodic oscillations in intensities and Doppler shifts, which have also been observed using the EUV Imaging Spectrometer on Hinode \citep[e.g.,][]{2009A&A...503L..25W,2009ApJ...696.1448W,2010ApJ...713..573M}, and their phase speed is close to the local sound speed \citep{2020A&A...638A...6S}. Slow waves are observed to have periods of several minutes to a few tens of minutes and they can be damped by physical processes such as thermal conduction, compressive viscosity, radiation, divergence of magnetic field, density stratification, etc. \citep[e.g.,][]{2002ApJ...580L..85O,2012SoPh..280..137A,2014ApJ...789..118K}.

The observed oscillations can be associated with the lower atmospheric oscillations or they could be in-situ oscillations. Many authors have investigated the propagation of waves from the photosphere to the TR, as well as the relationship between upper and lower region oscillations \citep[e.g.,][]{,2020JApA...41...18Z,2020A&A...634A..63K,2021SoPh..296..179C}. They interpreted these waves as upward propagating slow magneteoacoustic waves as they propagate along magnetic field lines and their phase speed between two heights is almost equal to the local sound speed.

In addition, some authors have found evidence of downward propagation waves. \cite{2003A&A...410..315R} investigated chromosphere and TR oscillations. They discovered that 5 min oscillations dominate in the chromosphere, whereas shorter period oscillations (2-3 min) dominate in the TR. They also found no oscillations in the corona and suggested this could be due to downward reflection occurring in the solar TR. \cite{2001A&A...371.1137B} have found significant power in chromospheric and transition lines at frequencies ranging from 4-8 min. They also found no significant power in the coronal line. The origin of these oscillations has been linked to the p-mode oscillations.

In addition to propagating waves, the standing slow waves are also observed in hot coronal loops. In hot flaring coronal loops, oscillations were observed in Doppler velocity during flaring by Solar Ultraviolet Measurements of Emitted Radiation (SUMER)/SOHO and Solar X-Ray Telescope (SXT)/Yokoh \citep{2002ApJ...574L.101W}. The oscillations lasted for a few ten minutes. It has been found that FeXIX and FeXXI emission lines (formation T > 6 MK) show a phase shift of $\pi/2$, which is the signature of a standing slow mode \citep[e.g.,][]{2011SSRv..158..397W,2015ApJ...807...98Y}. 

In \cite{2015ApJ...800..129M}, the sunspot oscillations in the chromosphere were investigated in conjunction with SJI filters 1400 \AA\ and 2796 \AA. In both the filters, the global period increases from sunspot center to the penumbra. They also found that apparent horizontal velocities decrease from 12 km~s$^{ - 1}$ in the umbra to about 4 km~s$^{ - 1}$ in the penumbra . On the basis of inclined field geometry, they have proposed that these oscillations are the signature of magnetoacoustic waves propagating upward. Since they have analyzed sunspot penumbra, they ought to consider the inclined geometry of the field lines and projection to estimate the properties of the slow waves \citep{2014A&A...561A..19Y}. However, in our present work, we have analyzed multiple locations in the broad bright and dark regions in the solar TR. We did not require to consider the geometry of an inclined magnetic field, in fact it can be considered as a top part of a wide slab piercing the solar TR and consisting of homogeneous magnetized plasma where the wave is supposed to be evolved and generating its signature in the TR layer.

\hfill \break
In the present work, we determined the physical properties of the detected oscillations by calculating the phase difference between intensity and velocity oscillations. We do not estimate the phase speed of the propagating wave (or oscillating antinodes in the case of standing mode) by traditional plase-lag/time-delay technique as adapted by many previous literatures \citep[e.g.,][]{2016A&A...585A.110K,2017ApJS..229...10J,2020A&A...634A..63K}. Instead, we use the well established theory of slow waves to draw the physical picture of the phase-relation between intensity and velocity in the solar TR.
Slow magnetoacoustic longitudinal oscillations are characterized by the propagation of density and velocity perturbations along the magnetic field. Because the solar TR is an optically thin and collisionally excited region, intensity ($I$) is proportional to the square of the density ($I\propto\rho^2$). This leads to the relationship between density perturbation $\rho_1$ and intensity perturbation $I_1$, i.e. $I_1/I_0$ = 2$\rho_1/\rho_0$ , from the liner approximation. As a result, intensity can be used as a proxy for density to interpret the type of the wave mode.

Many theoretical studies have predicted (e.g., based on forward modeling) that in the ideal MHD case a propagating wave mode manifests a phase difference of 0 and $\pm\pi$ between intensity and Doppler velocity, while a standing mode shows a phase difference of $\pm\pi/2$ \citep[e.g.,][]{1990A&A...236..509D,2010A&A...510A..41K,2010ApJ...721..744K,2013A&A...551A.137M}. \cite{2009ApJ...696.1448W} have reported the upward propagating slow magnetoacoustic wave on the basis of in-phase relationship between intensity and Doppler shift oscillations. Also, \cite{2003A&A...402L..17W,2003A&A...406.1105W} have reported slow mode standing wave in their analysis as there is a phase difference of $\pm\pi/2$ between intensity and velocity \citep[also see,][]{2008ApJ...681L..41M}. The present work shows that, in both the regions, the phase shift between intensity and Doppler velocity is within the range of -180\degree\ to +180\degree, according to wave phase analysis (bright and dark). Phase distribution peaks at -119\degree, 33\degree\ and 102\degree\ in the bright region. The centre peak is near to 0 and other two peaks are near to $\pm\pi/2$. We found the signature of a propagating wave as well as the signature of a standing wave in the bright region. In the dark region, phase distribution peaks at -153\degree, 6\degree\ and 150\degree. The central peak is near to 0, and other two peaks are near to $\pm\pi$. These phase difference values suggested both upward and downward propagating waves' signatures. 

We have also plotted the phase difference arrows on the cross power wavelet and wave coherence map (see, Figure \ref{fig:figure7}). In both regions (bright and dark), we found that different wave modes exist at different frequencies and times. For example, in bright region, downward propagating and standing waves are present at different frequencies. Similarly, in dark region, upward and downward propagating waves are present at different frequencies. Therefore, these wave modes (propagating and standing waves) are associated with different oscillations.

We found some level of consistency in explaining two of the observed phase peaks in both bright and dark regions when comparing with theoretical calculations of the 0.5D model for propagating slow magnetoacoustic modes (see, Sect. \ref{sec:phase_analysis_TR}). However, 1D model considering non-adiabatic effects, and a varying tube velocity with height (hence refractions) or more realistic 3D MHD models including the different damping mechanisms are needed in future to analyze the phase shifts of these oscillations. More recently the phenomenon of heating-cooling imbalance is being explored in a number of studies to model the solar atmosphere \citep[e.g.,][]{2021SoPh..296..105P,2022SoPh..297....5P} and it can be interesting to look into a 3D model including heating-cooling imbalance to calculate the phase shifts of magnetoacoustic oscillations.

We cannot directly infer the effect of the magnetic field on observations of various slow wave modes from our work, but we can comment on the energy flux carried by waves propagating in different regions (bright and dark). \cite{2009SoPh..258..219F} have investigated wave propagation from the photosphere to the corona numerically using hydrodynamic and magnetohydrodynamic modelling. According to the authors, the total energy transmitted to the corona in a magnetic case, (i.e. Thermal + kinetic + magnetic) is twice the amount of energy transmitted in a non magnetic case. Since the network region (bright patch) is surrounded by a strong magnetic field, propagating waves in this region may carry more energy than propagating waves in the inter-network region (dark patch), which is surrounded by a weak magnetic field.

\section{Acknowledgement}

We thankfully acknowledge the valuable remarks of referee that improved our manuscript. KS acknowledge CSIR for providing grant for her research work. We acknowledge the use of IRIS spectral data, and wavelet tool of Torrence and Compo (1998). IRIS is a NASA small explorer mission developed and operated by LMSAL with mission operations executed at NASA Ames Research Center and major contributions to downlink communications funded by ESA and the Norwegian Space Centre. A. K. Srivastava acknowledges the ISRO Project Grant (DS\_2B-13012(2)/26/2022-Sec.2) for the support of his research. The work of TJW was supported by NASA grants 80NSSC18K1131, 80NSSC18K0668, 80NSSC22K0755, 80NSSC21K1687 as well as the NASA Cooperative Agreement 80NSSC21M0180 to CUA. JJGA thanks to Investigadores por M\'exico-CONACYT (CONACYT Fellow), CONACYT LN 315829 (2021) and CONACYT-AEM 2017-01-292684 grants for partially support this work. The program "investigadores for M\'exico", project 1045 sponsor space Weather Service Mexico (SCIESMEX).

\section{Data Availability}

For spectroscopic study, we used observational data from IRIS that can be found at \url{https://www.lmsal.com/hek/hcr?cmd=view-recent-events&instrument=iris} and the data for the HMI on-board SDO are available in this at \url{http://jsoc.stanford.edu/ajax/lookdata.html}.







\appendix

\section{Spectra at dark region}
\label{spectra}

\begin{figure*}
  	\mbox{
   \centering
   	\includegraphics[width=.8\linewidth]{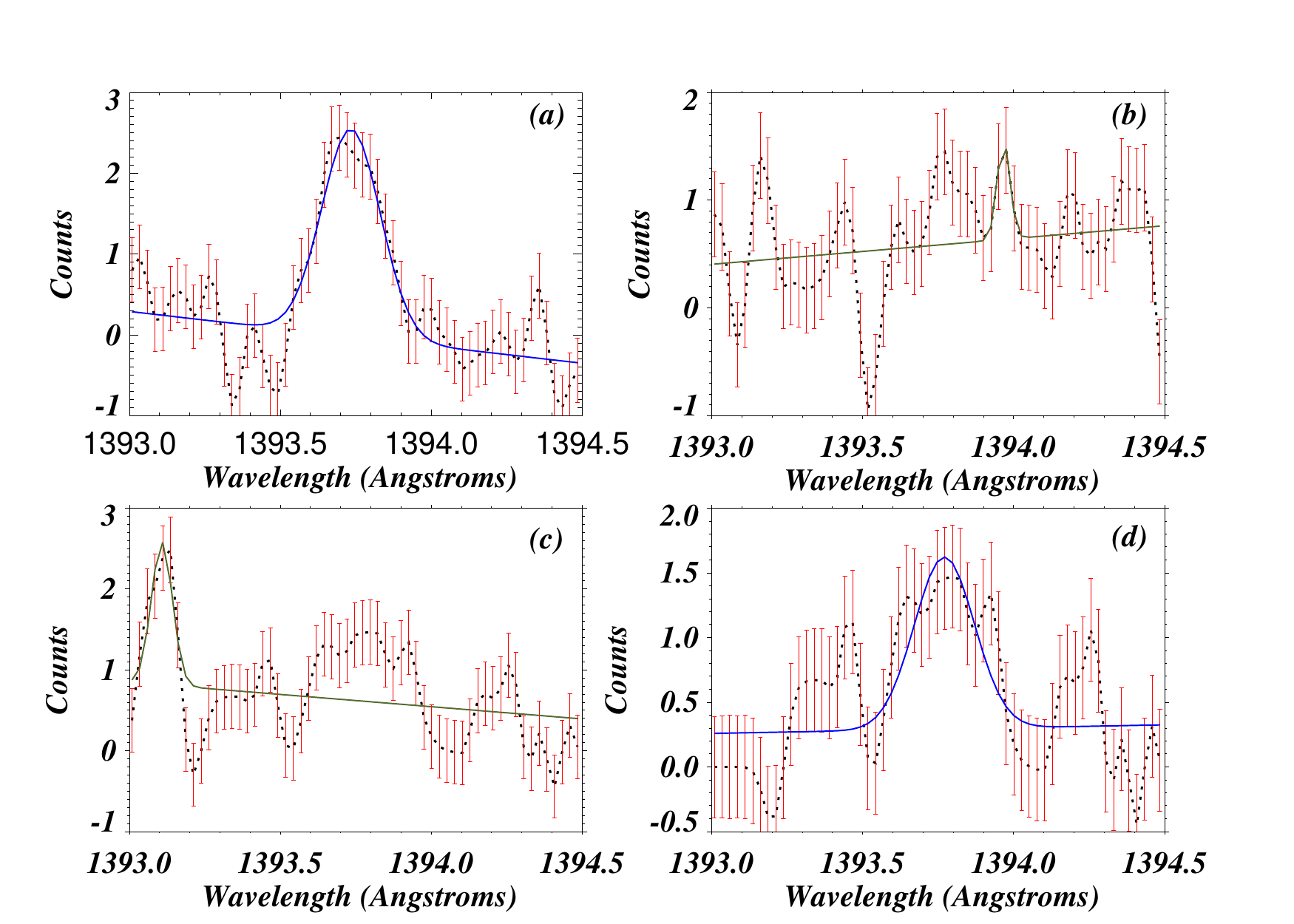} 
   	}
 \caption{ Spectral line profiles in different times in different locations. Figure (a) shows an example of a improved spectral line after binning. The spectral line is the black dotted curve, the fitted Gaussian curve is the blue line over plotted, and the red bars are the error bars. Figures (b) and (c) show an example of a failed fit. The auto fitted Gaussian curve, which gives the misleading spectral information, is shown in olive curve. After manually reducing the noise value, Figure (d) shows an example of a reliable fitted spectral line.}
  \label{fig:figure9}
\end{figure*}


Because the signal in the dark region is quite noisy, we used $4\times4$ binning (four in time and four in Y) to improve the signal to noise ratio. After binning, the signal to noise ratio is enhanced at most of the locations. However, the signal cannot be distinguished from the background at few points, which were excluded. Each spectra in the dark zone is carefully verified, and Gaussian fitting was performed to those pixels where the spectra was clearly visible. As a result, we have obtained the signal-induced intensity and velocity oscillations rather than background noise oscillations. We took one pixel and checked the spectral line profile at each time. We have seen three different types of spectra$\colon$

1. After binning, the spectra is improved, i.e. the signal is stronger than the noise, and hence the intensity and velocity signals can be estimated using a single Gaussian fitting. Figure \ref{fig:figure9} (a) shows an example of such spectra.
\hfill \break
2. Spectra in which the S/N has not yet been improved and the line (signal) cannot be distinguished from the background continua (noise). These spectra were not fitted instead, the intensity and velocity were interpolated later. An example of such spectra is shown in Figure \ref{fig:figure9} (b).
\hfill \break
3. Spectra with enhanced (S/N), although noise is stronger than the line (signal) at some other wavelengths (not in the proximity of 1393.77 \AA) in the specified wavelength range 1393.00 \AA\ – 1394.50 \AA. Such spectra can be seen in Figure \ref{fig:figure9} (c). The line is visible at wavelength 1393.77 \AA, but the routine fitted the noise instead of the line because it fits the peak of maximum intensity. Because the intensity in this example is at its maximum at 1393.10 \AA, the routine fitted the background (noise) rather than the spectral line (signal). We reduced the noise manually, and the routine fitted the signal peak rather than the noise. The fitting after noise reduction is shown in Figure \ref{fig:figure9} (d).

\hfill \break
Similarly, we checked spectrum profile at each Y location and estimated the intensity and velocity oscillations due to the signal but not due to the noise.

Moreover, we can also put some condition to the line width of the fitted profile to ensure the reliability of fitted profile. Instrumental broadening and thermal broadening are two components of line width measurements. For both the instrumental and thermal profiles, the measured Gaussian line width should (> $\sqrt{(instrumental\ width)^2 + (thermal\ width)^2}$ i.e., 0.056 \AA) \citep{1998ApJ...505..957C}. The line width for the fitted profiles shown in Figure \ref{fig:figure2} (d) and (f) are also small, so we checked the value of line width for these profiles and and value are larger than the 0.056 \AA. Similarly, we have checked the line width of all fitted spectral profiles and where it is smaller than 0.056 \AA\ we have flagged those values and interpolated later using the fitted profiles.

\section{Examples of cross power wavelet and wave coherence map in bright and dark regions}
\label{phase-arrows}

\begin{figure*}
  	\mbox{
   \centering
   	\includegraphics[width=.8\linewidth]{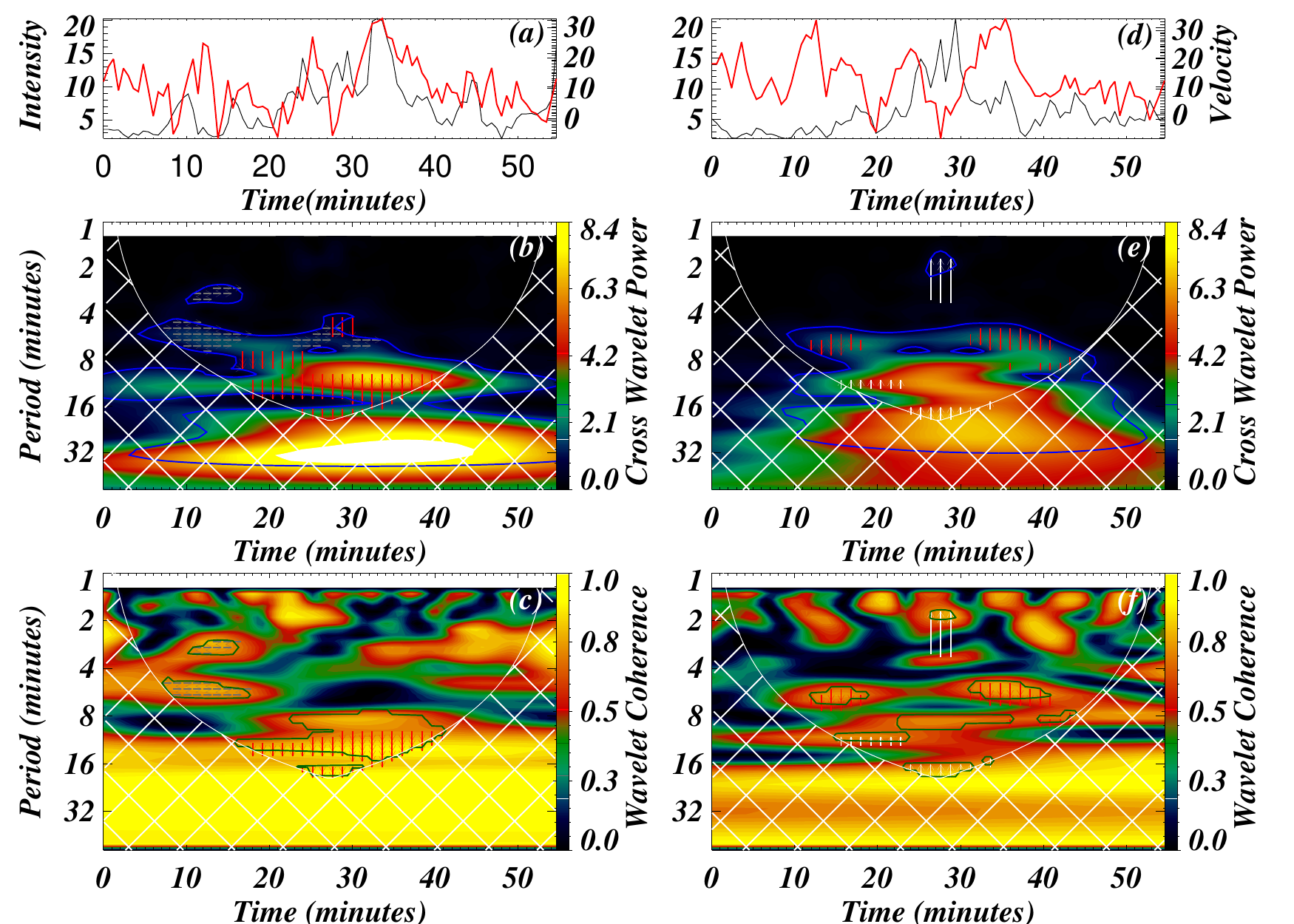} 
   	}
 \caption{The time-series of intensity (black) and Doppler velocity (red) at two different locations in the bright region are represented in panels (a) and (d). The cross power wavelet between two time series is shown in panels (b) and (e) and the wavelet coherence between two series is shown in panels (c) and (f). The other description of panels (a)-(f) are same as given in Figure \ref{fig:figure7}. }
  \label{fig:figure10}
\end{figure*}

\begin{figure*}
  	\mbox{
   \centering
   	\includegraphics[width=.8\linewidth]{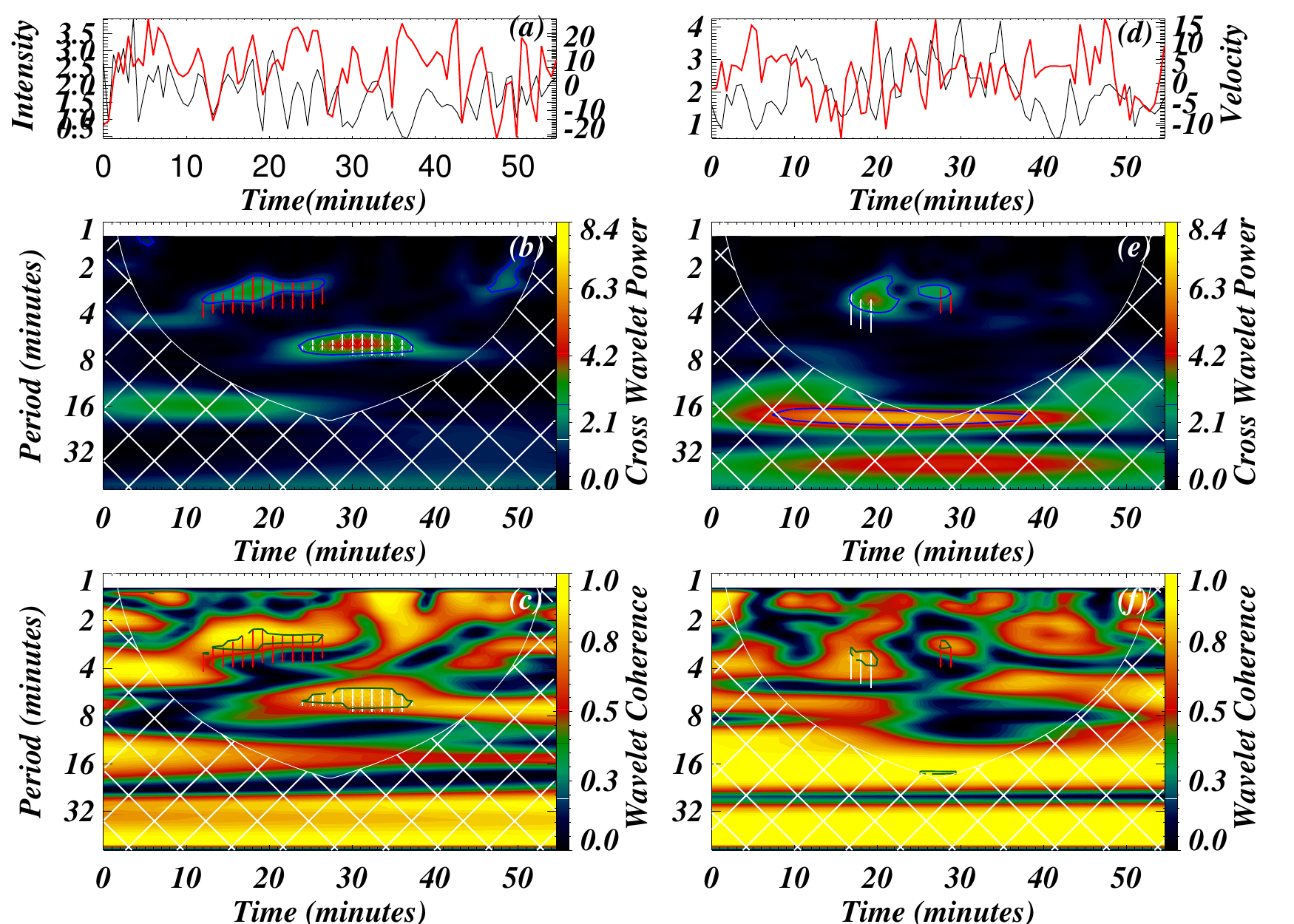} 
   	}
 \caption{The time-series of intensity (black) and Doppler velocity (red) at two different locations in the dark region are represented in panels (a) and (d). The cross power wavelet between two time series is shown in panels (b) and (e) and the wavelet coherence between two series is shown in panels (c) and (f). The other description of panels (a)-(f) are same as given in Figure \ref{fig:figure7}.}
  \label{fig:figure11}
\end{figure*}

In both bright and dark regions, we computed the phase difference between intensity and velocity time series. As shown in Figure \ref{fig:figure7}, we have drawn the phase difference arrow on both the cross power and wave coherence map for the chosen representative locations. On the cross power map, phase difference arrows are only plotted where cross power is significant, i.e., where cross power exceeds 95 \% global significance. The phase difference arrow is also plotted on the wave coherence map where cross power is significant along with wave coherence greater than 0.70. In Figure \ref{fig:figure7} (c), the cross power map shows the signature of both propagating waves and standing waves. A red arrow indicates downward propagating waves, while a gray arrow indicates standing waves. However, after applying a threshold for wave coherence, the red arrows become less prominent compared to the gray ones. Thus, for the bright region we have shown some other examples where both standing and downward propagating waves are clearly visible. The cross power wavelet and the wavelet coherence at two locations are shown in Figure \ref{fig:figure10} in this appendix. For both the locations, we have drawn arrows indicating phase differences. The presence of both propagating and standing waves can be seen in both locations at different frequencies. In a similar manner, we have shown cross power wavelet coherence for two different locations in the dark regions. An downward propagating wave is indicated by a red arrow, while a upward propagating wave is indicated by a white arrow (see, Figure \ref{fig:figure11}). Both examples show upward and downward propagating waves.

\label{lastpage}
\end{document}